\documentclass[times,final,longtitle]{elsarticle}

\usepackage{amssymb}
\usepackage{latexsym}

\usepackage{url}
\usepackage{tikz}
\usepackage[compat=1.1.0]{tikz-feynman}
\usepackage{xparse}
\RequirePackage{luatex85}

\NewDocumentCommand\semiloop{O{black}mmmO{}O{above}}
{%
\draw[#1] let \p1 = ($(#3)-(#2)$) in (#3) arc (#4:({#4+180}):({0.5*veclen(\x1,\y1)})node[midway, #6] {#5};)
}
\tikzset{
  photon/.style={decorate, line width=0.8pt, decoration={snake}, draw=black},
  fermion/.style={draw=black, line width=0.8pt, postaction={decorate}, decoration={markings, mark={at position .55 with {\arrow[scale=1.8,>=latex]{>}}}}}
}

\begin{document}


\begin{frontmatter}

\title{Searching in the dark: the hunt for the dark photon}%

\author[1]{Alessandra Filippi\corref{cor1}}%
\ead{filippi@to.infn.it}
  
\author[2]{Marzio De Napoli}



\address[1]{Istituto Nazionale di Fisica Nucleare sez. di Torino, via P. Giuria, 1, Torino 10125, Italy}
\address[2]{Istituto Nazionale di Fisica Nucleare sez. di Catania, Via S. Sofia, 62, Catania 92125, Italy}


\begin{abstract}
The existence of Dark Matter (DM) is a well established fact since many decades, thanks to the observation of the effects of its gravitational interaction with the ordinary matter in the Universe. However, our knowledge of the Dark Matter features is still rather scarce. Indeed, one of the biggest quests in fundamental science today is the investigation of Dark Matter nature, from its origin to its composition, and the way its constituents interact with the ordinary matter, apart from gravity. Huge and ambitious efforts have been spent in the last years into its identification, concentrating especially on the search of viable Weakly Interacting Massive Particle  candidates. However, no positive results have been achieved so far along this direction. On the other hand, many fascinating new ideas and models for its interpretation have been blooming: among them, an intriguing hypothesis is that the Dark Matter constituents could be neutral under Standard Model interactions, but they could interact through a new, still unknown, force under a ``hidden" charge. This new  hidden symmetry would be mediated by a massive gauge boson, the dark photon, which is expected to couple to the Standard Model via a kinetic mixing. 
The search for such a massive mediator has been pursued with large enthusiasm and dedication in the latest years,
as its observation could be within the reach of many already existing experimental facilities, both based on accelerators or in smaller scale setups. This report reviews the present status and progress of the experimental searches in this field.
\end{abstract}

\begin{keyword}
Dark photon\sep Dark matter\sep Experimental techniques for Dark Matter detection
\end{keyword}

\end{frontmatter}

\section{Introduction}
\label{intro}
\begin{figure}
\centering
\resizebox{6truecm}{!}{
\begin{tikzpicture}[node distance=2.5cm and 2.5cm]  
\begin{feynman}
\vertex (v1);
\vertex [right=of v1] (v2);
\vertex [right=of v2] (v3);
\vertex [right=of v3] (v4);
\draw[photon] (v1) -- (v2) node[midway,above=0.1cm] {$\gamma$};
\draw[photon] (v3) -- (v4) node[midway,above=0.1cm] {$A'$};
\semiloop[fermion]{v2}{v3}{0}[$\Psi'$];
\semiloop[fermion]{v3}{v2}{180}[$\Psi$][below];
\end{feynman}
\end{tikzpicture}
}
    \caption{Diagram showing the kinetic mixing of the SM photon with a dark photon $A^\prime$ at the one-loop level. $\Psi$ is any massive particle charged under both hypercharge U(1)$_Y$ and the secluded U(1)$_D$ symmetries.}
    \label{fig:kineticMixingGraph}
\end{figure}
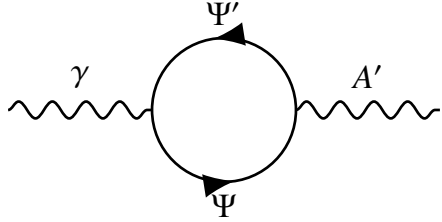

The gravitational interactions of Dark Matter (DM) provide the solid ground upon which the foundations of its existence are based. The first suggestion of its presence was proposed by Zwicky in the early 1930s: in the calculation of the dispersion velocity of the Coma cluster galaxies \cite{Zwicky33}, a large disagreement (of a factor about 400)
between the cluster's mass value deduced from the virial theorem and that
expected from the luminosities of its components was found. A way to explain such discrepancy was to postulate the existence a new form of non-visible matter, the ``{\it dunkel materie}" or Dark Matter. The idea had then been shelved for several decades, until it was recovered to explain the trend of the rotational velocity of galaxies, observed by Rubin and collaborators \cite{rubin80}: the fact that it resulted approximately flat instead of featuring the expected $1/\sqrt{r}$ behavior again called for the presence of a form on non-visible matter.
More recently, other gravitational effects were observed like those based on gravitational lensing \cite{gravlenses06}, on Bullet cluster \cite{bullet00} and on the cosmic microwave background measurements \cite{planckResults15}. They further fostered the hypothesis of a form of invisible massive matter which interacts gravitationally with the ordinary matter of the Universe, and possibly, but just feebly, in some other ways  (perhaps electroweakly, certainly not electromagnetically as it does not emit nor absorb electromagnetic radiation). 
According to the measurements performed by the WMAP \cite{wmap07} and Planck telescopes \cite{wmapAndPlanck}, it is nowadays ascertained that the DM constitutes at least the 85\% of the mass of our Universe.
While its existence is a consolidated fact, the issue of its composition and nature is, however, still open and this investigation represents indeed one of the main endeavours in physics today. Even though in the last decades large efforts were put both on the theoretical and on the experimental sides, on one hand to provide models which could describe its nature and the expected interaction patterns with the visible matter, and on the other one to investigate processes which could hint at the interaction mechanisms, at present a solid identification is  still far from reach. For this reason, many experimental approaches are presently being exploited, at high center-of-mass energy accelerators as well as in low-energy underground experiments: the variety of the possible phenomenology that could be involved is  wide and the parameters for the description of the related effects are basically unconstrained by theory in the absence, so far, of clean and driving experimental findings. It looks however clear that every observation of anomalies, especially on astrophysical scales, calls for some sort of new physics effect beyond the Standard Model of particle physics (SM), and can potentially have some sensitivity to DM existence.
A variety of DM candidates of various masses and subject to different kinds of interactions has been proposed over the years. 
Theoretically well-motivated candidates are represented by Weakly Interacting Massive Particles (WIMPs), assumed as cold thermal relics in thermal equilibrium with photons since the early Universe. On account of this  feature, their mass is expected in a wide, rather high range, from 10 GeV to 10 TeV, the lower bound being determined by the Lee-Weinberg cosmological limit \cite{LeeWeinberg77}. The WIMP paradigm has always been considered as one of the most solid for the DM identification, especially on account of the so-called ``WIMP miracle", the somewhat surprising match between the experimentally observed DM abundance and the value expected for it under the assumption of non-relativistic weakly interacting particles in thermal equilibrium with SM particles until the {\it freeze-out} \cite{miracle85}.
However, so far all WIMPs searches, both at colliders or based on direct observations, have been unsuccessful. 
To-date, the strongest bound to the WIMP-nucleon spin-independent elastic cross-section is posed by the XENON1T Experiment \cite{xenon1T18}, which excludes values down to $\sigma\sim 4\times 10^{-47}$ cm$^2$.
So, new different candidates are clearly needed.
From a thermodynamical point of view, the general accepted scenario is that  non-relativistic heavy DM is in thermal equilibrium with ordinary matter, a condition reached via subsequent annihilations proceeding through the exchange of intermediate gauge bosons, mediators of a new interaction which does not affect SM particles. However  this scenario, compatible with the WIMPs paradigm, can also hold if the DM particles are relativistic, with a much lighter mass but still consistent with the cosmological constraints from the Universe formation history --the lower limit being 3.3 keV \cite{lymanForest}. The light mass range can accommodate new candidates as, for instance, sterile neutrinos, expected around 10 keV \cite{zurek14, petraki13}.
The hypothesis of thermal light DM is an alternative appealing formulation also able to provide an explanation to the anomalous values (lower than expected) of the baryon temperature in the time lapse between 190 and 240 Myr since the Big Bang, as recently measured \cite{edge, lightDM}. 
Due to the negative results achieved so far in the search for WIMPs, a new wide campaign for the exploration of lighter forms of DM candidates, in the sub-GeV range or even lower, started recently. Of course, experimental searches in the tens of keV region  still represent enormous challenges at present, but at least the MeV level can be within reach for precise experiments in the near future. 
In the sub-GeV mass range, one of the simplest  hypotheses for Dark Matter particles identification is that it can belong to a ``hidden sector" secluded from the Standard Model, whose mutual interactions could be mediated by a massive gauge boson \cite{arkani09, pospelov09, hooper12}.
The concept of ``dark photon", also named as ``hidden photon" or ``heavy photon" and usually denoted as $A^\prime$ (or sometimes $U$), was introduced for the first time by Holdom \cite{holdom86}, who proposed  the possible existence of an additional spin-one gauge boson acting as the mediator of a further, hidden, U(1)$_D$ symmetry. This symmetry group kinetically mixes with the SM  hypercharge U(1)$_Y$. The coupling  of this additional gauge boson to the electric charge is expected to be suppressed, by a factor that is, however,  {\it a priori} unknown and covers about ten orders of magnitude, in the $10^{-12}-10^{-2}$ range. The kinetic mixing  between the dark and the SM photon would provide a ``portal" through which the hidden sector could be accessed, potentially allowing the properties of the  hidden particles belonging to it to be investigated. 
Fig. \ref{fig:kineticMixingGraph} shows in diagrammatic form how the kinetic mixing mechanism can occur assuming that a doublet of ``hidden" $\Psi\; (\Psi')$ DM particles exists, charged under both the SM hypercharge gauge group  and  the dark symmetry: the interaction between the dark and the SM photon is  realized at loop level, and the $\Psi$ ($\Psi'$) particles can have very large masses, even far above the SUSY-breaking scale.

The $A'$, however, is not the only possible candidate as hidden gauge mediator. Other portals are viable, that can be mediated by dark particles of different spins and which can possibly convey different sorts of interactions. While the dark photon has spin one and mediates the ``vector portal", a scalar dark CP-even or -odd Higgs  could mediate the so-called ``scalar portal", a dark fermionic mediator field (like a right-handed neutrino) could mediate the ``neutrino portal", while a pseudoscalar axion could mediate the ``axion portal". Of all these gauge groups have four dimensions and are renormalizable, with the exception of the non-renormalizable five-dimensional axion one \cite{darksectors16}. 
In principle any number of new gauge groups can be added to the SM, as long as its symmetry is not broken; therefore, many possible extensions of the minimal hidden sector model are possible and were indeed proposed, disclosing a variety of possible different experimental signatures and suggesting a larger number of new DM candidates. These scenarios are generally known as ``rich dark sectors".
This review aims at describing the present state-of-art in the search of a vector gauge boson as a conveyor  of  the interactions between ordinary and Dark Matter in a minimal secluded scenario.
The  paper is organized as follows. Sec. \ref{astrophysics} illustrates some  astrophysical
anomalies which could have a straightforward explanation if the DM is  charged under a new hidden symmetry; 
Sec. \ref{richSectors} shortly describes some of the most peculiar features of minimal hidden dark sectors extensions, which can provide some alternative explanations of the observed effects. 
A detailed description of the expected properties of dark photons will be accounted for in Sec. \ref{properties}. 
In particular, in Sec. \ref{theory} a short overview of the theoretical formalism for the description of an $A'$ and, more in general, of ``hidden" Dark Matter particles, as well as the expected interaction cross-sections, will be given. The production mechanism of $A'$s will be described in Sec. \ref{production}, and the expected decay patterns in Sec. \ref{decays}. 
Sec. \ref{experiments} will finally summarize the experimental approaches on which the current efforts in dark photon (and Dark Matter at large) searches are based. 
The techniques for the detection of dark photons through their decays will be described  with special focus on  experiments operating at low energy accelerators, in several different experimental environments. 
Results from searches in high energy $pp$ collisions as at the LHC (and specifically, at the ATLAS and CMS experiments)  will not be described in this review because of the extension of the topic, and the fact that in most of the cases a more complex $A'$ production through a scalar (via Higgs) or fermionic portal is implied, which goes beyond the vector portal scenario scope.

\section{Physics motivation}
Several anomalous effects have been observed over the years in terrestrial, astrophysical or cosmological experiments.
Some of them can be accounted for by introducing the concept of secluded sectors, while 
a few are still waiting for a more convincing interpretation, so for their thorough description more complex theoretical models are needed. 
Most of the models able to provide an explanation for these effects assume a relatively light form of Dark Matter (LDM), with candidates with masses of at most 10 GeV and on the order of 100 MeV for the dark mediator $A'$. In the following LDM particles, in the MeV--10 GeV mass range, will be generally denoted as $\chi$ (their CP properties being wherever disregarded, for the sake of notation simplicity).

\label{astrophysics}
\subsection{Astrophysical anomalies}
The recently observed astrophysical anomalies with a natural explanation in the hypothesis of secluded Dark Matter  include the excess of the positron cosmic ray flux, the extended gamma-rays spectrum emitted from the Galactic Center, and the observation of an anomalous monochromatic 3.56 keV X-ray line in the spectrum of several galaxy clusters. In the years, however, some of the current interpretations have been disproved or their range of validity has been reduced.
Early observations of a positron excess in cosmic ray fluxes, as compared to the electron fraction, were reported by several payload balloons experiments like HEAT \cite{heat95}, PPB-BETS \cite{ppb-bets08}, ATIC \cite{atic08}, CAPRICE \cite{caprice01} and PAMELA \cite{pamela09}.
Observations at an energy of about 10 GeV were more recently reported by the Fermi Gamma-Ray Telescope \cite{fermi12}, while  the Alpha Magnetic Spectrometer AMS-02 \cite{ams13}  provided an additional confirmation extending the observed positron spectrum energies up to 200 GeV. The observed positron excess calls for the existence of other positron sources in addition to the elementary interaction of cosmic ray nuclei with the interstellar medium, which for instance   provides the correct energy degrading description of the measured antiproton flux. One possibility is that the positron excess might derive from the direct annihilation of Dark Matter in lepton pairs, which however must occur at a  much larger rate than the typical Dark Matter thermal cross-section, $\langle \sigma v \rangle \simeq 3\times 10^{-26}\; \mathrm{cm}^{3}\mathrm{s}^{-1}$ \cite{thermalXsections09}. Nonetheless, if the DM annihilation occurs through the production of a dark photon, the cross-section can be subject to the so-called ``Sommerfeld enhancement" \cite{arkani09}, that can justify the rise of the observed positron yield at high energies as deriving from the dark photon direct decay in lepton pairs. The Sommerfeld enhancement increases the production rate mainly at lower velocities, keeping  the integrated thermal relic abundance within the correct limits. 
The most recent AMS-02 observations show, however, that the experimental positron spectrum is softer than predicted by light DM theories, and therefore this interpretation is acceptable only in the case of a heavy $A'$, with a mass in the 1--3 TeV interval, and with a DM annihilation cross-section spanning the  $\langle \sigma v \rangle \simeq (6-23)\times 10^{-24}\; \mathrm{cm}^{3}\mathrm{s}^{-1}$ range \cite{AMSXsectionsLimits13}. On account of this expectation, it looks like the unknown positron source is more likely to have a pulsar origin \cite{pulsar13}.
Lighter Dark Matter candidates are required, on the other hand, to explain the gamma-ray emission spectrum from the Galactic Center as observed by the Fermi Large Area Telescope \cite{gammaray1_11, gammaray2_11, gammaray3_11}. Dark matter annihilation to lepton or hadron
pairs is again one of the most reliable interpretations for the observed effect, as well as its annihilation to a dark photon and its subsequent decay to SM particles \cite{annihilationFermiLAT11, annihilationFermi09}.
Dark photons can also provide a direct  explanation of the 3.56 keV X-ray line observed in the emission spectrum of a number of galaxy clusters \cite{xrayline09}. The monochromatic X-ray line should come from the radiative decay of an excited Dark Matter particle $\chi^\ast$ ($\chi^\ast\rightarrow \chi\gamma$), produced in pairs through the DM self-rescattering  mediated by dark photons (``eXciting Dark Matter" model, XDM \cite{xdm16}). Further discussions on DM self-interaction mechanisms will be reported   in Sec. \ref{richSectors}.
A different important effect, whose interpretation could benefit from the  existence of a dark photon, is the issue concerning the measurement of the muon anomalous magnetic moment $a_\mu= (g_\mu-2)/2$. As it is nowadays well known, its measured value is in a $3.6\sigma$ tension with the value predicted by SM \cite{gMinus2_11}. The exchange of a dark photon via a triangular diagram could be responsible for such a disagreement.
The most recent experiments, though, have almost completely ruled out the parameters ranges favored by this interpretation, provided the dark photon decays exclusively to SM particles. On the other hand, there is still  some chance for this interpretation, with a coupling on the order of $\sim 10^{-4}$, in case the dark photons decay invisibly to LDM particles as well as in some extended versions of DM models.

\subsection{Extended hidden sectors}
\label{richSectors}
More complex extensions have been proposed, most of which aim to overtake the picture of collisionless cold DM \cite{planckResults15} suggesting alternative mechanisms and non-thermal regimes, to provide different interpretations for other anomalous signals which cannot find a direct explanation in the minimal secluded scenario picture. As a consequence new effects are also expected, whose future possible observation would contribute to steer the search along some new and more effective directions; indeed, one of the challenges of future experiments will be to find some solid hints of ``exotic signatures".
Most of these extended models can be tailored to provide, for instance, reasonable alternative explanations to  the mentioned muon magnetic moment issue (which can revive under new hypotheses), to the proton charge radius puzzle \cite{protonradius} and to the recent observation of unexpected effects in nuclear transitions, performed by the ATOMKI Collaboration \cite{berylliumtransitions, helium4_x17}. 
This is a particularly interesting case, in which the existence of a new $J_\pi = 1^+$ vector boson, supposedly mediating a new, fifth fundamental force with some coupling to SM particles \cite{X17_Theo1,X17_Theo2}, can explain the two significant enhancements observed in the angular correlation of $e^+e^-$ pairs emitted in the transitions to the ground state of the $^8$Be 18.15 MeV level ($J_\pi = 1^+$) and of the $^4$He 21.01 MeV one ($J_\pi = 0^-$, $M0$ forbidden transition). Constraints on such a new particle, however, require an extension of the minimal vector portal scenario such to include a protophobic coupling to Up and Down quarks.
One of the most critical issues calling for further developments and extensions is the tension between the observed DM halo profiles \cite{haloprofiles94, haloprofilesFlores} and those obtained from simulations in the hypothesis of cold DM \cite{tensionHaloprofiles12, tensionHaloprofiles15}.  A way to reconcile such a conflict is to assume that Dark Matter can self-interact via the exchange of a light mediator. The models allowing for DM self-interaction do not alter the interpretation of the mentioned astrophysical anomalies, but they permit much tinier values for the coupling among the mediator and the SM fields. A consequence of self-interacting DM is the possible formation of bound states formed by DM particles, which can annihilate and therefore provide an alternative source for DM production, overstepping the mentioned Sommerfeld enhancement mechanism \cite{boundstates09, boundstates10}. If existing, their observation could be within reach at high luminosity colliders in the next future. The possibility of self-interactions would also admit many-body scattering processes and interactions mediated by the strong force (among the so-called SIMPs, Strongly Interacting Massive Particles \cite{simps}). This could move the focus to non-thermal scenarios in which the relationship between masses and couplings could be substantially different from the currently accepted thermal paradigm.
The inelastic Dark Matter (iDM) scenarios are, in turn, based on the hypothesis that the elastic interaction of DM with SM is suppressed. The simplest models assume that the lightest stable DM species, $\chi_1$, can only interact inelastically with SM through the exchange of a vector mediator leading to a slightly heavier $\chi_2$ DM state
\cite{tucker05,izaguirre-2017}. In this way the relic DM abundance is dominated by a $\chi_1\chi_2\to\mathrm{SM}$ inelastic process, known as {\it coannihilation} \cite{coannihilation}. The mass difference $\Delta$ between the two DM states is on the order of $\sim 100$ keV, the typical halo WIMP kinetic energy. More details on iDM interaction mechanisms will be given in Sec. \ref{decays} and \ref{BDinvisible}, for their relevance for potential DM discovery.

\section{Dark Photon expected properties, production and decays}
\label{properties}
\subsection{Dark Photon parameters: coupling and mass}
\label{theory}
To introduce the fundamental parameters for the description of a dark photon, the coupling and its mass, let us consider now in more detail a minimal secluded model, in which the 
the dark sector is represented by just a single extra U(1)$_D$ gauge group. In the hypothesis of a vector mediator, the gauge Lagrangian may be written in the following form:
\begin{equation}
    \label{eq:lagrangian}
    \mathcal{L}_{gauge} = 
    -\frac{1}{4}B_{\mu\nu}B^{\mu\nu}
    -\frac{1}{4}F^\prime_{\mu\nu}F^{\prime\mu\nu} +
    \frac{1}{2}\frac{\epsilon}{\cos\theta_W}F^{\prime}_{\mu\nu}B^{\mu\nu}
\end{equation}
where $B_\mu$ and $A^\prime_\mu$ are the mediator field of the SM U(1)$_Y$ symmetry and the dark U(1)$_D$ gauge group, $B_{\mu\nu} = \partial_\mu B_\nu -\partial_\nu B_\mu$ is the field strength tensor of U(1)$_Y$, respectively,  
$F^\prime_{\mu\nu} = \partial_\mu A^\prime_\nu -\partial_\nu A^\prime_\mu$ is the corresponding one of U(1)$_D$, $\theta_W$ the weak mixing angle and $\epsilon$  the kinetic mixing parameter. The dominant mixing between sectors can be assumed to involve SM photons only, as the possible mixing with the heavy $Z$ boson, at next-to-leading order, is negligible being further suppressed by a 1/$m^2_Z$ factor.
The dimensionless parameter $\epsilon$ determines the magnitude of the coupling between $A^\prime$ and the SM sector; as mentioned in the introduction, the mixing through loop diagrams (Fig. \ref{fig:kineticMixingGraph}) makes this constant small.
The mixing term in Eq. (\ref{eq:lagrangian}) can be removed by the redefinition of the SM hypercharge field 
$B_\mu \rightarrow B_\mu + \epsilon A^\prime_\mu$, so that the coupling of the dark photon to the ordinary electromagnetic current emerges as:
\begin{equation}
     \mathcal{L}_{dark,\gamma} = -e\epsilon A'_\mu J^\mu_{em}. 
     \label{eq:Lint}
\end{equation}

Integrating out the fields in the loop diagram of Fig. \ref{fig:kineticMixingGraph} one gets for $\epsilon$ the following expression:
\begin{equation}
    \epsilon \sim \frac{g_Y g_D}{16\pi^2}\log\left(\frac{m_\Psi}{m_\Psi^\prime}\right) \sim 10^{-3}-10^{-1}.
\end{equation}
where $g_Y$ is the SM electroweak coupling constant  (being $\alpha = g^2_Y/(4\pi)\simeq 1/137$  the QED fine-structure constant) and $g_D$ the dark coupling constant between the $A'$ and the DM particles in the hidden sector.  
In some cases the one-loop contributions vanish (like in large extended GUT-inspired groups), so two-loops diagrams play the dominant contribution:  the coupling in this case is reduced to the $10^{-5}-10^{-3}$ range \cite{holdom86, epsilonOneLoop09, epsilonOneLoop08, epsilonOneLoop10, epsilonOneLoop2_09}. According to some string  based models, values as small as $\sim10^{-12}$ can also be expected \cite{epsilonStrings09, epsilonStrings10, epsilonStrings11}. So, in general, the value of the kinetic mixing can vary is a wide range, and there is no {\it a priori} constraint from the theory which limit it.
The same is valid for the second parameter of the theory, the dark photon mass $m_{A^\prime}$. As anticipated in the introduction, the focus of the most recent efforts has been set mainly on the investigation of the range between 1 MeV and 10 GeV, the lower bound being fixed by the existing observations at accelerators and by astrophysical or cosmological constraints, 
while the upper one is determined by the maximum reach of experiments at high-intensity  colliders. Higher mass values, and also extra gauge bosons suggested by extended models, can be probed by the present high-energy facilities, such as the LHC.

\subsection{Dark Photons production}
\label{production}
\subsubsection{Dark Photons from Dark Matter annihilation}
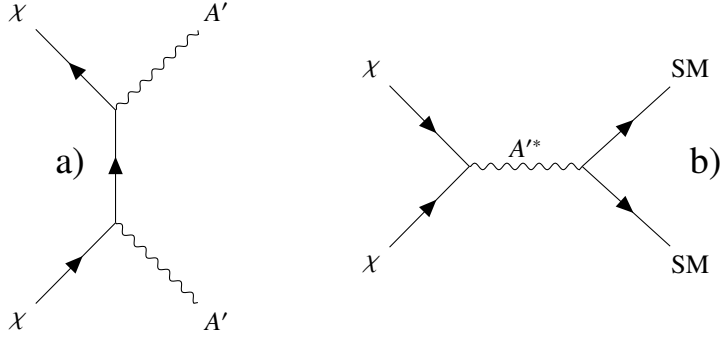
\begin{figure}
\centering
\begin{minipage}{.45\textwidth}
\centering
\begin{tikzpicture}[scale = 5]
    \begin{feynman}
    \vertex (b);
    \vertex [above left=of b] (a) {$\chi$};
    \vertex [above right=of b] (e) {$A'$};
    \vertex [below=of b] (c);
    \vertex [below left=of c] (d) {$\chi$};
    \vertex [below right=of c] (f) {$A'$};
    \diagram* {
      (b) -- [fermion] (a),
      (b) -- [boson] (e),
      (d) -- [fermion] (c) -- [boson] (f),
      (c) -- [fermion] (b),
    };
  \end{feynman} 
\end{tikzpicture}
\end{minipage}
\begin{minipage}{.45\textwidth}
\centering
\begin{tikzpicture}
  \begin{feynman}
    \vertex (b);
    \vertex [below left=of b] (a) {$\chi$};
    \vertex [above left=of b] (e) {$\chi$};
    \vertex [right=of b] (c);
    \vertex [below right=of c] (d) {SM};
    \vertex [above right=of c] (f) {SM};
    \diagram* {
      (a) -- [fermion] (b),
      (e) -- [fermion] (b),
      (c) -- [fermion] (d),
      (c) -- [fermion] (f),
      (b) -- [boson, edge label=$A^{\prime\ast}$] (c),
    };
  \end{feynman}

\end{tikzpicture}
\end{minipage}
\put(-260,0){{\Large a)}}
\put(-20,0){{\Large b)}}
\caption{Diagrams for the production of a dark photon $A'$  via annihilation: $\chi\chi\rightarrow A'A'$ (a) or  $s$-channel resonant production into SM particles $\chi\chi\rightarrow A^{\prime\ast}\rightarrow\mathrm{SM\;SM}$ (b).}
\label{fig:annihilation}
\end{figure}

Dark photons can be produced by DM particles via annihilation processes according to  two modes, depending on the relative mass between $A^\prime$ and $\chi$: if $m_{A^\prime} < 2m_{\chi}$, the annihilation occurs via a ``secluded" process like $\chi\chi \rightarrow A^\prime A^\prime$ (see Fig. \ref{fig:annihilation}a);  conversely, when $m_{A^\prime} > 2m_{\chi}$,  via the  $\chi\chi\rightarrow A^{\prime\ast} \rightarrow \mathrm{SM}\;\mathrm{ SM}$ ``direct annihilation", where the virtual $A^{\prime\ast}$ is exchanged in the $s$-channel and SM particles in the final state are  produced 
(see Fig. \ref{fig:annihilation}b).

In the case of secluded DM annihilations, the production rate scales as \cite{secludedAnnihilations}
\begin{equation}
    \langle \sigma v\rangle \sim \frac{g_D^4}{m^2_\chi}
\end{equation}
so it only depends on the dark $g_D$ coupling and on the $m_\chi$ mass of the DM particles, but not on the $\gamma$-$A^\prime$ mixing $e\epsilon$ parameter -- which means that the process would be hard to be detected by a particle physics experiment. 
In the case of direct DM annihilation the rate scales according to
\begin{equation}
    \langle\sigma v\rangle \sim \frac{g^2_D\alpha\epsilon^2 m^2_\chi}{m^4_{A^\prime}}
\end{equation}
{\it i.e.} it is inversely dependent of the 4-th power of the dark photon mass. Since in this case the $m_\chi/m_{A^\prime}$ mass ratio and the $g_D$ dark coupling constant can at most be $\mathcal{O}(1)$, limits can be set to $\epsilon$ to preserve the consistency with the minimum expected thermal annihilation rate:
\begin{equation}
    \frac{\alpha g^2_D\epsilon^2}{4\pi}\left(\frac{m_\chi}{m_{A^{\prime}}}\right)^4 \gtrsim \langle \sigma v\rangle_{relic} m^2_\chi.
\end{equation}

\subsubsection{Dark Photons from Standard Model particles interactions: electron Bremsstrahlung}
\label{QEDbackgrounds}

\begin{figure}[!ht]
\begin{minipage}[b]{.35\textwidth}
\begin{center}
\begin{tikzpicture}[scale = 5]
    \begin{feynman}
    \vertex (a) {$e^-$};
    \vertex [right=of a] (b);
    \vertex [right=of b] (f);
    \vertex [below=of b] (c);
    \vertex [left=of c] (d) {$Z$};
    \vertex [right=of c] (m);
    \vertex [right=of m] (e) {$Z$};
    \vertex [above right=of f] (g); 
    \vertex [below right=of f] (l) {};
    \vertex [above right=of g] (h) {$e^- (\ell^-,\; \chi)$};
    \vertex [below right=of g] (i) {$e^+ (\ell^+,\; \chi)$};
    \diagram* {
      (a) -- [fermion] (b),
      (b) -- [fermion, edge label=$e^-$] (f),
      (b) -- [boson, edge label=$\gamma$] (c), 
      (d) -- [fermion, very thick] (c) -- [fermion, very thick] (e),
      (f) -- [boson, edge label=$A^{\prime}$] (g),
      (f) -- [fermion, edge label=$e^-$] (l),
      (i) -- [fermion] (g),
      (g) -- [fermion] (h)
    };
  \end{feynman} 
\end{tikzpicture}
\end{center}
\end{minipage}
\hfil
\begin{minipage}[b]{.25\textwidth}
\begin{center}
\begin{tikzpicture}
  \begin{feynman}
    \vertex (a) {$e^-$};
    \vertex [right=of a] (b);
    \vertex [above right=of b] (c);
    \vertex [below right=of b] (f);
    \vertex [below left=of f] (h);
    \vertex [left=of h] (i) {$Z$};
    \vertex [right=of h] (m);
    \vertex [right=of m] (l) {$Z$};
    \vertex [right=of f] (g) {$e^-$};
    \vertex [above right=of c] (d) {$e^- (\ell^-,\; \chi)$};
    \vertex [below right=of c] (e) {$e^+ (\ell^+,\; \chi)$};
    \diagram* {
      (a) -- [fermion] (b),
      (b) -- [fermion, edge label=$e^-$] (f),
      (i) -- [fermion, very thick] (h) -- [fermion, very thick] (l),
      (b) -- [boson, edge label=$A^{\prime}$] (c),
      (h) -- [boson, edge label=$\gamma$] (f),
      (f) -- [fermion] (g),
      (e) -- [fermion] (c),
      (c) -- [fermion] (d),
    };
  \end{feynman}

\end{tikzpicture}
\end{center}

\end{minipage}
\put(-250,120){{\Large a)}}
\put(-20,120){{\Large b)}}

\caption{Diagrams showing the radiation of an $A^\prime$ from an electron in a dark photon-bremsstrahlung process. The dark photon can be radiated from an electron in the initial state (a) or as final state radiation (b); in both cases it is on shell and can travel some distance before decaying in $e^+e^-$, two leptons ($\ell^+\ell^-$) or two generic $\chi$ DM particles.}
\label{fig:bremsstrahlung}
\end{figure}
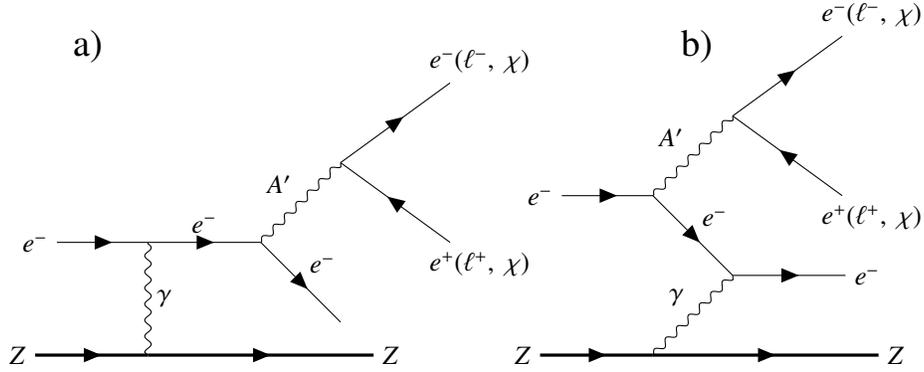

High luminosity fixed targets experiments may exploit the mechanism of radiative emission of dark photons by electron beams incident on a high $Z$ atomic number target. The process, also known as $A'$-strahlung for short, is depicted in Fig. \ref{fig:bremsstrahlung} and is analogous to a SM photon bremsstrahlung, even though the different coupling and mass of the $A^\prime$ lead to different kinematics and rates.

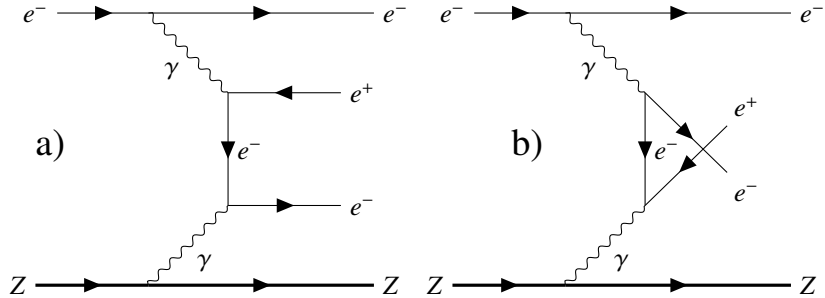
\begin{figure}[!ht]
\centering
\begin{minipage}{.45\textwidth}
\centering
\begin{tikzpicture}[scale = 5]
    \begin{feynman}
    \vertex (a) {$e^-$};
    \vertex [right=of a] (b);
    \vertex [right=of b] (k);
    \vertex [right=of k] (c) {$e^-$};
    \vertex [below right=of b] (d);
    \vertex [right=of d] (e) {$e^+$};
    \vertex [below=of d] (f);
    \vertex [right=of f] (g) {$e^-$}; 
    \vertex [below left=of f] (h);
    \vertex [right=of h] (j);
    \vertex [right=of j] (l) {$Z$};
    \vertex [left=of h] (i) {$Z$};
    \diagram* {
      (a) -- [fermion] (b) -- [fermion] (c),
      (d) -- [boson, edge label=$\gamma$] (b),
      (d) -- [fermion, edge label=$e^-$] (f), 
      (e) -- [fermion] (d),
      (f) -- [fermion] (g),
      (f) -- [boson, edge label=$\gamma$] (h),
      (i) -- [fermion, very thick] (h) -- [fermion, very thick] (l)
    };
  \end{feynman} 
\end{tikzpicture}
\end{minipage}
\begin{minipage}{.45\textwidth}
\centering
\begin{tikzpicture}
  \begin{feynman}
    \vertex (a) {$e^-$};
    \vertex [right=of a] (b);
    \vertex [right=of b] (k);
    \vertex [right=of k] (c) {$e^-$};
    \vertex [below right=of b] (d);
    \vertex [below right=of d] (e) {$e^-$};
    \vertex [below=of d] (g);
    \vertex [above right=of g] (f) {$e^+$}; 
    \vertex [below left=of g] (h);
    \vertex [right=of h] (j);
    \vertex [right=of j] (l) {$Z$};
    \vertex [left=of h] (i) {$Z$};
    \diagram* {
      (a) -- [fermion] (b) -- [fermion] (c),
      (d) -- [boson, edge label=$\gamma$] (b),
      (d) -- [fermion, edge label=$e^-$] (g), 
      (d) -- [fermion] (e),
      (f) -- [fermion] (g),
      (g) -- [boson, edge label=$\gamma$] (h),
      (i) -- [fermion, very thick] (h) -- [fermion, very thick] (l)
    };
  \end{feynman} 
  
\end{tikzpicture}
\end{minipage}
\put(-300,0){{\Large a)}}
\put(-120,0){{\Large b)}}

\caption{Diagrams (a, b) for the Bethe-Heitler (space-like) trident reactions.}
\label{fig:QEDdiagrams_BH}
\end{figure}

\begin{figure}[!ht]
\begin{minipage}[b]{.35\textwidth}
\begin{center}
\begin{tikzpicture}[scale = 5]
    \begin{feynman}
    \vertex (a) {$e^-$};
    \vertex [right=of a] (b);
    \vertex [right=of b] (f);
    \vertex [below=of b] (c);
    \vertex [left=of c] (d) {$Z$};
    \vertex [right=of c] (k);
    \vertex [right=of k] (e) {$Z$};
    \vertex [above right=of f] (g); 
    \vertex [below right=of f] (l) {};
    \vertex [above right=of g] (h) {$e^-$};
    \vertex [below right=of g] (i) {$e^+$};
    \diagram* {
      (a) -- [fermion] (b),
      (b) -- [fermion, edge label=$e^-$] (f),
      (b) -- [boson, edge label=$\gamma$] (c), 
      (d) -- [fermion, very thick] (c),
      (c) -- [fermion, very thick] (e),
      (f) -- [boson, edge label=$\gamma$] (g),
      (f) -- [fermion, edge label=$e^-$] (l),
      (i) -- [fermion] (g),
      (g) -- [fermion] (h)
    };
  \end{feynman} 
\end{tikzpicture}
\end{center}
\end{minipage}%
\hfil
\begin{minipage}[b]{.25\textwidth}
\begin{center}
\begin{tikzpicture}
  \begin{feynman}
    \vertex (a) {$e^-$};
    \vertex [right=of a] (b);
    \vertex [above right=of b] (c);
    \vertex [below right=of b] (f);
    \vertex [below left=of f] (h);
    \vertex [left=of h] (i) {$Z$};
    \vertex [right=of h] (k);
    \vertex [right=of k] (l) {$Z$};
    \vertex [right=of f] (g) {$e^-$};
    \vertex [above right=of c] (d) {$e^-$};
    \vertex [below right=of c] (e) {$e^+$};
    \diagram* {
      (a) -- [fermion] (b),
      (b) -- [fermion, edge label=$e^-$] (f),
      (i) -- [fermion, very thick] (h) -- [fermion, very thick] (l),
      (b) -- [boson, edge label=$\gamma$] (c),
      (h) -- [boson, edge label=$\gamma$] (f),
      (f) -- [fermion] (g),
      (e) -- [fermion] (c),
      (c) -- [fermion] (d),
    };
  \end{feynman}
  
\end{tikzpicture}
\end{center}
\end{minipage}
\put(-250,120){{\Large a)}}
\put(-70,120){{\Large b)}}

\caption{Diagrams (a, b) for the QED radiative (time-like) trident reactions.}
\label{fig:QEDdiagrams_tri}
\end{figure}
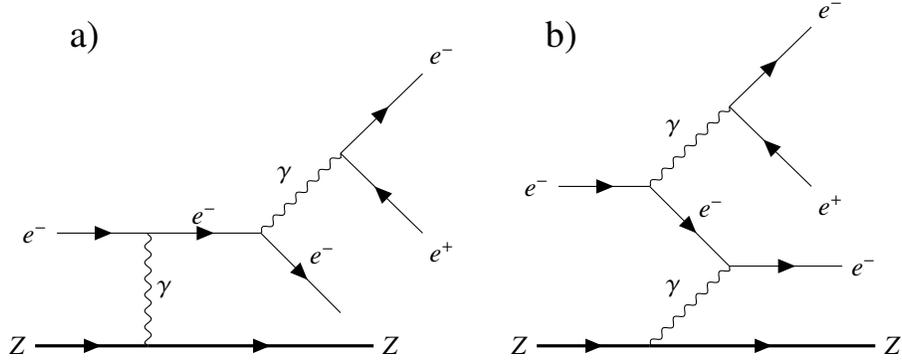

The energy-angle distribution of the dark photon radiated by an electron can be described using the Weiszsacker-Williams approximation \cite{weiszsackerWilliams86, weiszsackerWilliams86, weiszsackerWilliams74, weiszsackerWilliams77erratum, weiszsackerWilliams73}. According to this model, the scattering is treated as a Compton-like process and the resulting differential cross-section for the $e Z\rightarrow e^\prime A^\prime Z'$ reaction is given by \cite{epsilonOneLoop09}:
\begin{equation}
    \frac{d\sigma(e Z\rightarrow e' A' Z)}{dx\; d\cos\theta_{A^\prime}} = \frac{8\alpha^3\epsilon^2 E^2_0 x\sqrt{1-m^2_{A^\prime}/E^2_0}}{U^2}\Phi\left[\left(1-x+\frac{x^2}{2}\right)-\frac{(1-x)^2 m^2_{A^\prime}}{U^2}\left(m^2_{A^\prime}-\frac{Ux}{1-x}\right)\right]
    \label{eq:bremseq}
\end{equation}
$E_0$ being the energy of the incident electron, $\theta_{A^\prime}$ the opening angle of the radiated $A^\prime$, $x = E_{A^\prime}/E_0$ the fraction of the electron energy carried away by the dark photon. $U$ is a function related to the virtuality of the intermediate electron:
\begin{equation}
    U(x, \theta_{A^\prime}) = E^2_0\,x\,\theta^2_{A^\prime} + m^2_{A^\prime}\frac{1-x}{x}+m^2_e x
\end{equation}
with $m_e$ the electron mass. In Eq. (\ref{eq:bremseq}) $\Phi$ represents the effective dark photon flux, that is defined through the elastic  $G_{2,el}$ and the inelastic  $G_{2,in}$ form factors of the target nucleus, which parameterize the effects of the electron screening and the of size of the nucleus, via the formula:
\begin{equation}
    \Phi = \int_{t_{min}}^{t_{max}} \left(G_{2,el}(t)+G_{2,in}(t)\right)\frac{t-t_{min}}{t^2} dt
\end{equation}
where the integral limits are given, respectively, by $t_{min} = (m^2_{A^\prime}/2E_0)^2$ and $t_{max} = m^2_{A^\prime}$. Assuming that $m_e \ll m_{A^\prime}$ one can integrate Eq. (\ref{eq:bremseq}) over the angles and obtain the following cross-section:
\begin{equation}
    \frac{d\sigma}{dx} = \frac{8\alpha^3\epsilon^2\sqrt{1-m^2_{A^\prime}/E^2_0}}{m^2_{A^\prime}(1-x)/x + m^2_e x}\left(1-x+\frac{x^2}{3}\right)\Phi
    \label{eq:rate}
\end{equation}
which reduces to the ordinary photon bremsstrahlung cross-section when $m_{A^\prime}\rightarrow 0$. From Eq. (\ref{eq:rate}) one can deduce that the production rate of dark photons is proportional to ${\alpha^2\epsilon^2}/{m^2_{A^\prime}}$, thus suppressed by a factor ${\epsilon^2m^2_e}/{m^2_{A^\prime}}$ relative to photon bremsstrahlung. The $\Phi$ effective dark photon flux brings an overall suppression effect for large $A^\prime$ masses and small beam energies; the $U(x,0)$ function gets its minimum for $x \sim 1$, so that the most favorable condition is when the forward emitted $A^\prime$ carries away most of the beam energy. The emission angle $\theta_{A^\prime}$ has a limit value that is much smaller than the opening angle of the decay products of the dark photon, proportional to $m_{A^\prime}/E_0$. 
Competing reactions to the $A^\prime$-strahlung are the  QED Bethe-Heitler and radiative trident processes, depicted in the diagrams of Fig. \ref{fig:QEDdiagrams_BH} and Fig. \ref{fig:QEDdiagrams_tri}, respectively. The former is the dominant process, but its kinematics is different from the $A^\prime$ case so the separation of the two reactions is, in principle, possible. In fact, while the $A^\prime$ decays products are highly forward boosted and the recoiling electron is soft and scatters  at large angles, just one of the leptons from the Bethe-Heitler process is boosted, being the other much softer. Moreover, at higher pair energies the $A'$-strahlung cross-section is enhanced, while the Bethe-Heitler one is not.   

\subsubsection{Dark Photons from Standard Model particles interactions: $e^+e^-$ annihilation on bound electrons in atoms}
\label{annihilation_boundE}
\begin{figure}
\centering

\begin{minipage}{.45\textwidth}
\centering
\begin{tikzpicture}
  \begin{feynman}
    \vertex (b);
    \vertex [below left=of b] (a) {$e^-$};
    \vertex [above left=of b] (e) {$e^+$};
    \vertex [right=of b] (c);
    \diagram* {
      (a) -- [fermion] (b),
      (e) -- [fermion] (b),
      (b) -- [boson, edge label=$A^{\prime}$] (c),
    };
  \end{feynman}
\end{tikzpicture}
\end{minipage}
\begin{minipage}{.45\textwidth}
\centering
\begin{tikzpicture}[scale = 5]
    \begin{feynman}
    \vertex (b);
    \vertex [above left=of b] (a) {$e^+$};
    \vertex [above right=of b] (e) {$A'$};
    \vertex [below=of b] (c);
    \vertex [below left=of c] (d) {$e^-$};
    \vertex [below right=of c] (f) {$\gamma$};
    \diagram* {
      (b) -- [fermion] (a),
      (b) -- [boson] (e),
      (d) -- [fermion] (c) -- [boson] (f),
      (c) -- [fermion] (b),
    };
  \end{feynman} 
\end{tikzpicture}
\end{minipage}
\put(-300,0){{\Large a)}}
\put(-100,0){{\Large b)}}

\caption{Diagrams for the production of a Dark Photon $A'$  via resonant (a) and non-resonant (b) $e^+e^-$ annihilation.}
\label{fig:annihilation_ee}
\end{figure}
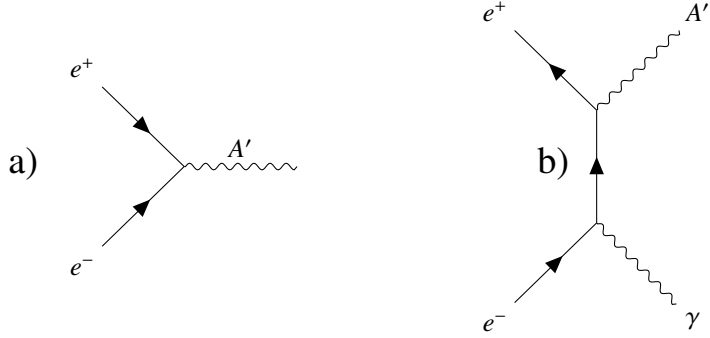

The diagrams of Fig. \ref{fig:annihilation_ee} describe the $A'$ production through  $e^- e^+\rightarrow A'$ resonant (a) and  $e^- e^+\rightarrow A'\gamma$ non-resonant (b) positron-electron annihilation. 
At a center-of-mass energy $\sqrt{s}$ of a few tens of MeV the two-photon production is  the dominant process in $e^+e^-$ annihilation on a bound electron; it is however possible for an  
$A'$ to replace one of the two SM photons.

The cross-sections for the annihilation processes scale as $\epsilon^2\alpha$ (resonant) or $\epsilon^2\alpha^2$ (non-resonant), to be compared to the $\epsilon^2\alpha^3$ dependence of the $A'$-strahlung process (see Eq. (\ref{eq:bremseq})). 
Specifically, the total cross-section for the resonant diagram (Fig. \ref{fig:annihilation_ee}a)  is \cite{resonant_annihilation}:

\begin{equation}
    \sigma_{res} = 
    \sigma_{peak}
    \frac{\Gamma^2_{A'}/4}{(\sqrt{s}-m_{A'})^2+\Gamma^2_{A'}/4} 
\end{equation}
where  $\sigma_{peak}=12\pi/m^2_{A'}$ is the resonant cross-section at $m_{A'}$, and $\Gamma_{A'}=\frac{1}{2}m_{A'}\epsilon^2\alpha$ is the $A'$ decay width in the  $m_{e}/m_{A'}\rightarrow 0$ limit;
this expression is valid in the  narrow width approximation, given that $\Gamma_{A'}/m_{A'}\ll1$ since $\epsilon \ll 1$.

The differential and total cross-sections for the non-resonant process shown in Fig. \ref{fig:annihilation_ee}b) are given by, respectively \cite{Marsicano1}:

\begin{equation}
    \frac{d\sigma_{nores}}{dz} = 
    \frac{4\pi\epsilon^2\alpha^2}{s}\left(
    \frac{s-m^2_{A'}}{2s}
    \frac{1+z^2}{1-\beta^2z^2}+
    \frac{2m^2_{A'}}{s-m^2_{A'}}
    \frac{1}{1-\beta^2z^2}
    \right)
\end{equation}

\begin{equation}
    \sigma_{nores}=
    \frac{8\pi\epsilon^2\alpha^2}{s}
    \left[
    \left(
    \frac{s-m^2_{A'}}{2s}
    \frac{m^2_{A'}}{s-m^2_{A'}}
    \right)
    \log\frac{s}{m^2_{e}}-
    \frac{s-m^2_{A'}}{2s}
    \right]
    \label{eq:annihilation_angdist}
\end{equation}
being $z$ the $A'$ emission angle cosine in the $e^+e^-$ rest frame, measured with respect to the positron direction, and $\beta=\sqrt{1-4m^2_{e}/s}$.

On the other hand, in the limit of a high  $E_+$ energy positron beam, the differential cross-section in the reference system of the laboratory as a function of the $A'$ photon energy, $y=E_{\gamma}^{lab}/(E_+ + m_e)$, is \cite{VEPP3}: 

\begin{equation}
    \frac{d\sigma}{dy}\sim\epsilon^2
    \frac{\pi\;r^2_{e}}{y\gamma^+}
    \left[
    \frac{(1+\mu)^2}{1-(y+\mu)}
    -2y
    \right]
\end{equation}
where $\mu=m_{A'}^2/s$ and $\gamma^+ = E_+/\mu \gg 1$,  while $r_e$ is the classical electron radius; the photon energy is limited by $y<(1-\mu)$. 

In the case of resonant positron annihilation the kinematics of the produced $A'$ is strongly constrained by the one-body nature of the final state: a dark photon with mass $m_{A'}$ is produced with energy $E_{res}=m_{A'}^2/(2m_{e})$, in the same direction of the incoming positron. For the non-resonant case, the $A'$ angular distribution in the center-of-mass frame, given by Eq. (\ref{eq:annihilation_angdist}), is boosted along the center-of-mass direction, due to the $1-\beta^2z^2$ factor. This results in a strongly forward peaked angular distribution, the  more the larger the $A'$ mass values. The limit value for the $A'$ emission angle in the lab is 
$\theta_{A'}^{max}\simeq{s-m_{A'}^2}/{(2m_{A'}E_{0})}$. The corresponding energy distribution ranges from $E_{res}$ (value at $m_{A'}$ peak) to the primary positron energy $E_{0}$, with an average value ${E_0}/2\cdot (1+{m_{A'}^2}/{(2m_{e}E_0)})$.

\subsubsection{Dark Photons from Standard Model particles interactions: Compton-like scattering}
\label{comptonlike}
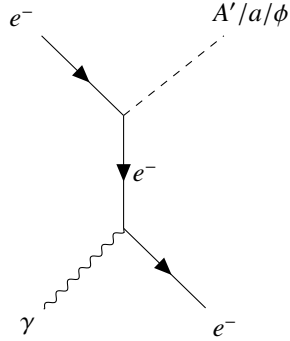
\begin{figure}
\centering
\begin{minipage}{.5\textwidth}
\centering
\begin{tikzpicture}[scale = 5]
    \begin{feynman}
    \vertex (b);
    \vertex [above left=of b] (a) {$e^-$};
    \vertex [above right=of b] (e) {$A'/a/\phi$};
    \vertex [below=of b] (c);
    \vertex [below left=of c] (d) {$\gamma$};
    \vertex [below right=of c] (f) {$e^-$};
    \diagram* {
      (a) -- [fermion] (b),
      (b) -- [fermion, edge label=$e^-$] (c),
      (d) -- [boson] (c) -- [fermion] (f),
      (b) -- [scalar] (e),
    };
  \end{feynman} 
\end{tikzpicture}
\end{minipage}

\caption{Diagram for the production of a Dark Photon $A'$ (or axion $a$ or dark scalar mediator $\phi$) via a Compton-like process induced by a real photon.}
\label{fig:comptonFig}

\end{figure}

An alternative method for Dark Photons production, studied in a quantitive way just recently \cite{compton}, is based on Compton-like emission in reactions induced by photons on bound electrons in the atoms of the target: $\gamma e^-\rightarrow A'e^-$ (see Fig. \ref{fig:comptonFig}). This reaction can be effective also for the production of axion-like pseudoscalar particles ($a$ in  Fig. \ref{fig:comptonFig}) or dark scalar mediators ($\phi$).

The cross-section for a Compton-like process is approximately given by
\begin{equation}
    \sigma_{Comp} \sim \frac{4\pi\alpha^2\epsilon^2}{s}\log \left(\frac{1-x^2_M}{x_m}\right)
\end{equation}
where $x_m$ ($x_M$) is the fraction of the $\sqrt{s}$ center-of-mass energy  corresponding to the mass of the electron $m_e = x_m\sqrt{s}$ (dark photon, $m_{A'} = x_M\sqrt{s}$). Numerically, for a 10 GeV photon beam impinging on a liquid hydrogen target and producing a 10 MeV $A'$ one gets for the total cross-section, approximately
\begin{equation}
    \sigma_{Comp} \sim 1.4\; \mathrm{pb} \left(\frac{\epsilon}{10^{-4}}\right)^2\left(\frac{0.1\;\mathrm{GeV}}{\sqrt{s}}\right)^2.  
\end{equation}
If the incident photon beam energy is much larger than the energy threshold for the production of an $A'$, the cross-section is practically independent on the mass of $A'$.
The differential cross-sections is forward peaked, as in the $A'$-strahlung case. As in all processes involving bound electrons in the target atoms, the atomic binding energy as well as screening effects and radiative corrections must be properly taken into account; in general, for a Compton-like process the cross-section scales with the target atomic number $Z$, while for brehmsstrahlung in nuclei it scales with $Z^2$: this means that the Compton-like process can be in principle favored versus resonant $A'$-strahlung by positrons, or $A'$-strahlung in nuclei when a proton beam is used.

\subsection{Dark Photon decay modes and branching fractions}
\label{decays}
\begin{figure}[!ht]
\centering
\includegraphics[scale=.15]{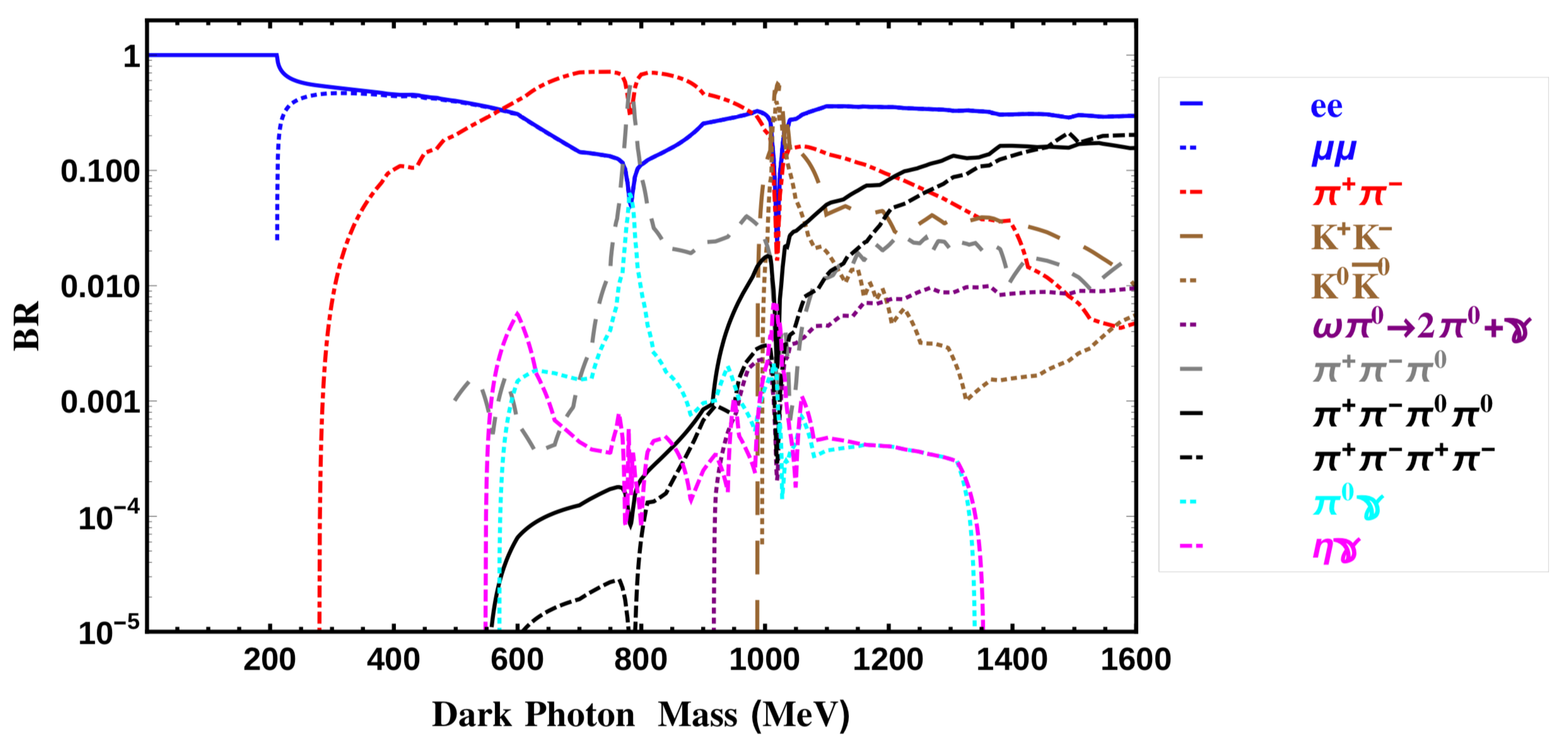}
\caption{Branching fractions for dark photons as a function of their mass (from Ref. \cite{branchingratios}).}
\label{fig:DPbranchingratio}
\end{figure}

According to the relative values of the masses of the hidden gauge mediator and of the particles belonging to the hidden sectors, the dark photon can undergo visible or invisible decays. 
If the mass of the $A'$ is lighter than twice the mass of any Dark Matter particle, the decays in DM is kinematically forbidden and the dark photon can only decay to SM particles: these are called ``{\it visible decays}".
A sketch of the trend as a function of the $A'$ mass of the branching fractions for the $A'$ visible decay in several channels is reported in Fig. \ref{fig:DPbranchingratio} (from Ref. \cite{branchingratios}).

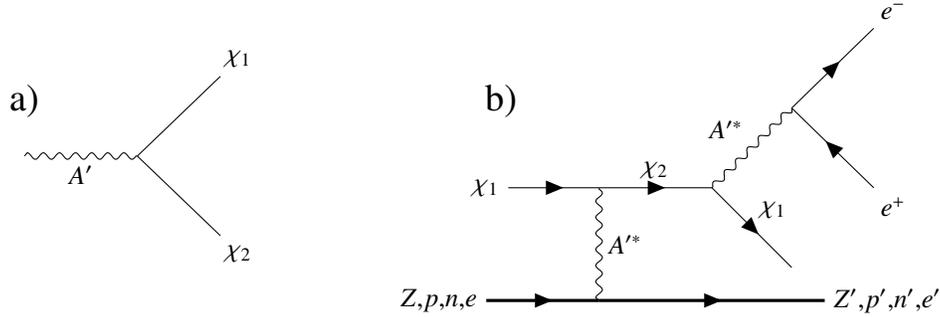
\begin{figure}
\begin{minipage}{.25\textwidth}
\centering
\begin{tikzpicture}
  \begin{feynman}
    \vertex (b);
    \vertex [below right=of b] (a) {$\chi_2$};
    \vertex [above right=of b] (e) {$\chi_1$};
    \vertex [left=of b] (c);
    \diagram* {
      (a) -- (b) [fermion],
      (e) -- (b) [fermion],
      (b) -- [boson, edge label=$A^{\prime}$] (c),
    };
  \end{feynman}
\end{tikzpicture}
\end{minipage}
\hfil
\begin{minipage}{.45\textwidth}
\centering
\begin{tikzpicture}[scale = 5]
   \begin{feynman}
    \vertex (a) {$\chi_1$};
    \vertex [right=of a] (b);
    \vertex [right=of b] (f);
    \vertex [below=of b] (c);
    \vertex [left=of c] (d) {$Z$,$p$,$n$,$e$};
    \vertex [right=of c] (k);
    \vertex [right=of k] (e) {$Z'$,$p'$,$n'$,$e'$};
    \vertex [above right=of f] (g); 
    \vertex [below right=of f] (l) {};
    \vertex [above right=of g] (h) {$e^-$};
    \vertex [below right=of g] (i) {$e^+$};
    \diagram* {
      (a) -- [fermion] (b),
      (b) -- [fermion, edge label=$\chi_2$] (f),
      (b) -- [boson, edge label=$A^{\prime *}$] (c), 
      (d) -- [fermion, very thick] (c),
      (c) -- [fermion, very thick] (e),
      (f) -- [boson, edge label=$A^{\prime *}$] (g),
      (f) -- [fermion, edge label=$\chi_1$] (l),
      (i) -- [fermion] (g),
      (g) -- [fermion] (h)
    };
  \end{feynman} 
\end{tikzpicture}
\end{minipage}
\put(-300,20){{\Large a)}}
\put(-120,20){{\Large b)}}

\caption{(a) Inelastic DM scenario: the $A'$ decays to a pair of different mass eigenstates $\chi_{1,2}$.  The unstable $\chi_2$ decays in flight, so the DM flux in a detector tens of meters away from the production point, as in a beam-dump experiment, is dominated by $\chi_1$ states only. (b) The $\chi_1$ can then upscatter off electrons, nucleons and/or nuclei to produce a new $\chi_2$ state which promptly de-excites delivering a three-body final state $\chi_{2} \rightarrow \chi_{1}e^{+}e^{-}$ through the dilepton decay of a virtual $A'$. }
\label{fig:iDM}
\end{figure}

The proper lifetime for a visible $A^\prime$ decay in SM particles is given by
\begin{equation}
\label{eq:decay}
    c\tau = \frac{1}{\Gamma} = \frac{3}{N_{eff}m_{A^\prime}\alpha\epsilon^2} \sim \frac{80\;\mathrm{\mu m}}{N_{eff}}\left(\frac{10^{-4}}{\epsilon}\right)^2\left(\frac{100\;\mathrm{MeV}}{m_{A^\prime}}\right)
\end{equation}
where $N_{eff}$ is the number of available decay channels ($N_{eff} = 1$ for $m_{A^\prime} < 2 m_\mu$, while $N_{eff} = 2+ R(m_{A^\prime})$ for $m_{A^\prime} \geq 2 m_\mu$, $R$ being the ratio between the $e^+e^-$ cross-sections for hadrons and dimuon productions). 
$c\tau$ represents the impact parameter for the detection of displaced vertices due to a dark photon decay, a typical value being around 80 $\mu m$ for dark photons of mass on the order of 100 MeV. From Eq. (\ref{eq:decay}) the inverse dependence of the decay length to $\epsilon^2$ appears, whose direct consequence is the possibility of investigating small couplings just by experiments able to detect considerably detached decay vertices, as will be described in detail in Sec. \ref{experiments}. 
On the other hand, if the $A'$ decays to other light dark particles of the hidden sector is kinematically allowed, ``{\it invisible decays}" are possible and, though being reduced in strength by a factor $\alpha_D^2$, they would uniformly suppress the branching fractions for decays in visible channels. The decays in DM particles escape of course the detection by particle detectors, but can be inferred by missing-mass or missing momentum techniques, as will be discussed in a  following Section. $A^\prime$ can also, in principle, decay into mixed final states containing both SM and dark particles. In this case the decay identification would benefit from missing-energy techniques, that are mostly insensitive to the set of particles produced in the final state.  
In iDM scenarios (see Sec. \ref{richSectors}) the Dark Photon couples to a pseudo-DM Dirac fermion $\chi$:  due to a spontaneous symmetry breaking in the dark sector, the $A'$ acquires a mass, and the $\chi$ splits into two Majorana mass eigenstates $\chi_1$ and $\chi_2$ with slightly different masses, see Fig. \ref{fig:iDM}a) \cite{izaguirre-2017}. The heavier DM particle $\chi_2$ is unstable and decays in-flight immediately, for instance through the $\chi_2\to\chi_1 A^{\prime\ast}\to\chi_1 e^+e^-$ channel. The interaction between the DM and SM particles occurs via $A'$ exchange through the inelastic (up)scattering $\chi_1 T\to \chi_2 T'$ depicted in Fig. \ref{fig:iDM}b), where $T (T')$ can be SM electrons, nucleons or even nuclei.
As will be described in Sec. \ref{BDinvisible}, beam-dump experiments are potentially able to detect the (up)scattering of the long-lived $\chi_1$ component with particularly striking signatures.

\begin{figure}[!ht]
\centering
\includegraphics[width=\linewidth]{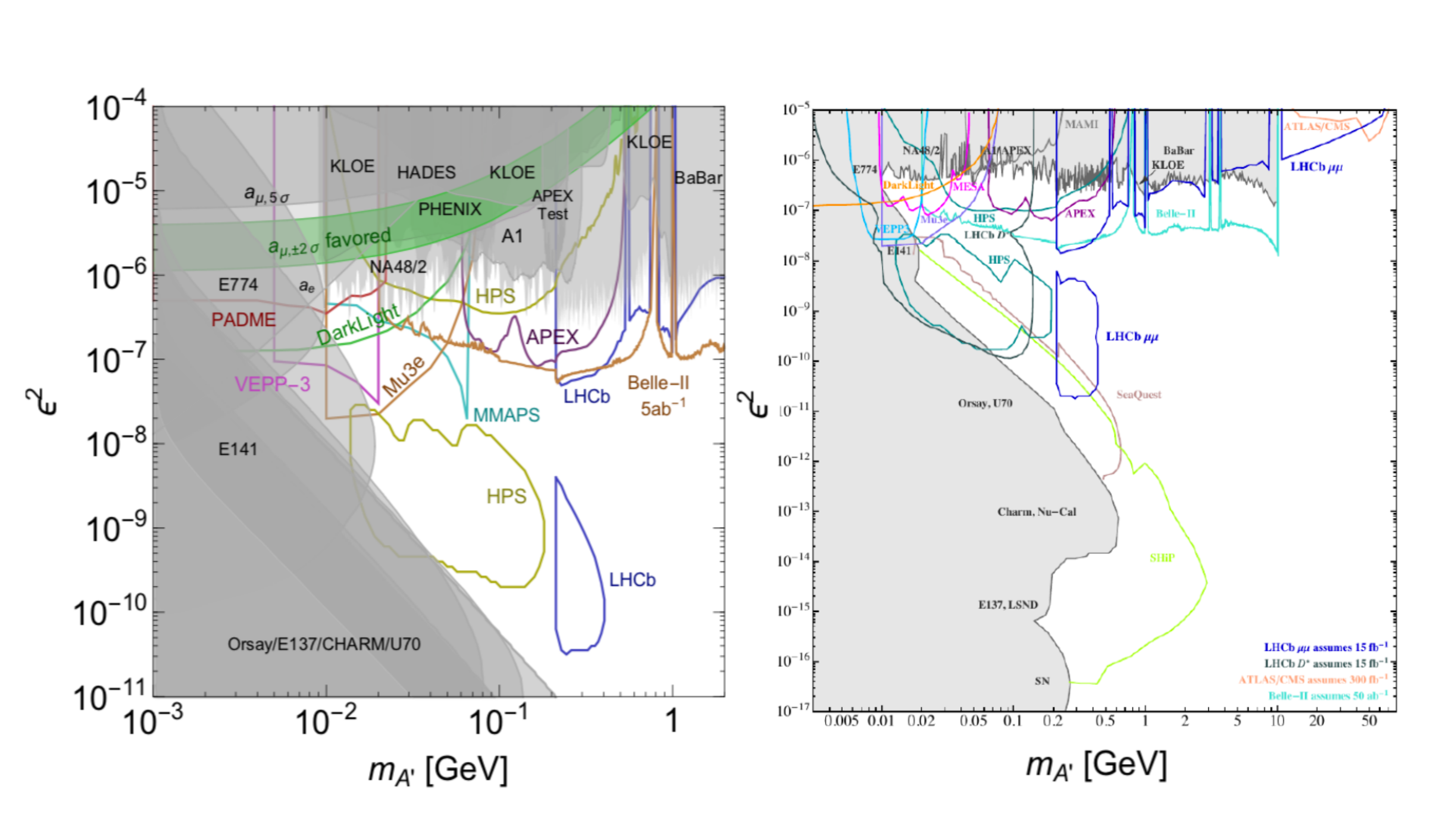}
\put(-200,75){{\Large a)}}
\put(-40,75){{\Large b)}}
\caption{Existing bounds (shaded regions) and projected sensitivities of ongoing/proposed experiments (lines) for dark photon in visible decays in the mixing strength {\it vs} $A'$ mass parameter plane (from
Refs.\cite{US-cosmic-vision,choi16,LHCb-2}): a) $\epsilon^2\; vs\; m_{A'}$ in a zoomed  parameter region; b) $\epsilon^2\; vs\; m_{A'}$ in a wider region covered also by some high energy collider experiments. Constrains are derived at 90\% C.L.
The bound marked as ``SN'' in the bottom-left part of the right panel comes from the astrophysical observations of the SN1987A supernova cooling \cite{SN1987A}.
The constraints from the measurement of the anomalous magnetic moment of the electron $a_{e}$ \cite{gMinus2DP09} and the preferred region to explain the discrepancy of the muon anomalous magnetic moment $a_{\mu}$ are also visible on the left panel.}
\label{fig:exclusionLimits}
\end{figure}

\section{Dark Photons and Light Dark Matter searches at accelerators}
\label{experiments}
In this Section we summarize the main features and strategies of past, present, and future experiments searching for the dark photon and Dark Matter candidates with mass lighter than 10 GeV at accelerator facilities. The focus will be set to experiments currently running (denoted in the following  with the superscript $^\circ$) or in the preparation or proposal stage (denoted as $^\ast$) at low energy accelerators; already concluded experiments will be denoted by the $^\dagger$ superscript  \footnote{list updated as of end of 2019 (circa)}. As already mentioned, a full description of the searches at high-energy colliders will not be reported here: a thorough review of the experimental strategies may be found, for instance, in Ref. \cite{LHC}, while  the latest results from CMS and ATLAS are reported in Refs. \cite{CMS,CMS_new} and \cite{ATLAS}, respectively.
Some ATLAS and CMS search outlooks at HL-LHC can be found in Refs. \cite{ATLAS_HL-LHC,CMS_HL-LHC}, and a review of some future experimental scenarios at the CERN accelerator complex is reported in Ref. \cite{CERN-Report}. 
For the extension of the topic, results from Direct Detection searches will not be discussed as well in this review  (see for example Ref. \cite{US-cosmic-vision} for a summary of new trends and findings in this investigation).
With respect to Direct Detection searches, experiments at accelerators offer significant advantages. They are independent of astrophysical uncertainties and all rely on the relativistic production and detection of hidden sector particles, thereby minimizing Lorentz dependent effects of various SM/DM mutual interactions. The direct detection sensitivity is, in fact, strongly subject to the mediator spin and/or the velocity dependence of interaction rates, whereas this does not hold in
accelerator-based experiments (where $v \sim c$).
In addition, the relativistic production of DM at accelerators allows to probe iDM scenarios (see Sec. \ref{richSectors}) not accessible to Direct Detection experiments, in which the up(scattering) of non-relativistic halo DM particles into heavier states is kinematically forbidden if the mass splitting is $\sim 100$ keV or larger \cite{izaguirre-2017}.
Finally, in experiments at accelerators the time of production and, in some cases, the direction of the DM ``beam'' driven by the incoming SM particles flux is known, contrary to direct searches.\\ 
The design of all the accelerator-based experiments for the search of a  vector boson is steered by two main goals: to measure the features of $A'$ production and to efficiently reject the SM background. This general target is pursued employing many different techniques which exploit  various $A'$ production modes and decay channels. The many possible experimental approaches are highly complementary, each with its own strengths and weaknesses. They differ under several respects, from the used beam  (electrons, positrons, hadrons) to its energy (from hundreds MeV to hundreds GeV), to the employed target (which can be thin, thick as in beam-dump setups or also missing, as in reactions at colliders), to the search strategy (counting experiments, bump-hunting, displaced vertex reconstruction, missing mass/energy/momentum  techniques). Many experiments have been explicitly designed to probe dark sectors, many others are used for hidden particle searches just as a by-product, often in fresh reanalyses of old data; nonetheless they are useful to derive  additional constraints on the hidden sector parameter space.\\ 
For the sake of simplicity, the experimental techniques described in the following paragraphs will be ascribed to the two wide categories of  \textit{Visible} and \textit{Invisible} decay searches introduced in Sec. \ref{decays}; experiments will be grouped according to their main research strategy: \textit{Bump hunting}, \textit{Beam dump} and \textit{Decay-vertex reconstruction} for visible decays, \textit{Missing mass}, \textit{Missing energy/momentum}, and \textit {Beam dump} for invisible decays searches. \\   
All the constraints on the parameter space reported in this paper will be shown as exclusion limits, derived at 90\% Confidence Level. More exclusion limits obtained assuming different parameters and/or DM models are discussed, for instance, in Refs. \cite{darksectors16,US-cosmic-vision,CERN-Report}. 
Note that some of the existing constraints have been extracted  from results of old experiments related to other searches, often with different event topologies (which could have a drawback in the detection efficiency and acceptance). Therefore, some caution is needed when considering such limits and the resulting extension of the excluded regions, since the recast of old results could suffer from a lack of experimental and analysis information and be affected by a sizeable systematic uncertainty.\\

\subsection{Visible decay searches}
As already mentioned in Sec. \ref{decays} if $m_{A'}<2m_{\chi}$ the $A'$ can decay only into SM particles. In this case the hidden mediator is usually searched through its leptonic decay $A'  \rightarrow e^+e^-,\; \mu^+\mu^-$, whose branching fractions are the largest ones irrespective of the $A^\prime$ mass and its production mechanism (as shown in Fig. \ref{fig:DPbranchingratio}). The exploited production mechanisms in such searches are the annihilation process 
($e^+e^- \rightarrow \gamma A'$), the bremsstrahlung of an $A^\prime$ by an electron ($e^- Z \rightarrow e^- Z A'$), Compton-like scattering induced by real photons ($\gamma e^-\rightarrow A' e^-$), peculiar meson decays like 
Dalitz'  ($\pi^0/\eta/\eta' \rightarrow \gamma A'$) or  rarest decays  (such as $K \rightarrow \pi A'$, $\phi \rightarrow \eta A'$, and $D^* \rightarrow D^0 A'$), and Drell-Yan reactions $q\bar{q} \rightarrow A'\rightarrow \ell^+\ell^-$ or $h^+h^-$ (with $h$  a generic hadron),  being these particularly effective in hadron colliders and proton fixed-target experiments.
Several constraints on the dark photon parameter space have been already set for the minimal scenario, and many others are foreseen  in the near future (see Fig. \ref{fig:exclusionLimits} (a,b) for an updated summary). Visible decays are mostly constrained from searches for di-electron or di-muon resonances \cite{BaBar-visible,Na48-2015,A1} and from the re-interpretation of data from fixed target or neutrino experiments in the low (sub-GeV) mass region \cite{E141,E137-SLAC,E774-Fermilab}. 
In general, three regions can be identified in Fig. \ref{fig:exclusionLimits}, each  covered resorting to a different detection strategy. The upper part, corresponding to  $\epsilon^{2} \gtrsim 10^{-6}$ in the 10 MeV--10 GeV mass region for $A'$, is related to bump-hunt searches performed at beauty and kaon factories, electroproduction experiments and hadron induced reactions, and to the $a_\mu$ exclusion region: NA48/2$^\dagger$ (CERN) \cite{Na48-2015}, A1$^\dagger$ (MAMI-Mainz) \cite{A1} and BABAR$^\dagger$ (SLAC) \cite{BaBar-visible}  set the strongest bounds for $\epsilon$ ($> 10^{-3}$) in this mass range. 
The nearly triangular bottom-left shaded region $(10^{-11} \lesssim \epsilon^2 \lesssim 10^{-6},\; 1\;\mathrm{MeV}\lesssim m_{A'} \lesssim 200\;\mathrm{MeV}$) is covered by old beam-dump experiments such as E141$^\dagger$ (SLAC) \cite{E141}, E137$^\dagger$ (SLAC) \cite{Marsicano1, E137-SLAC, E137-SLAC-new}, E774$^\dagger$ (Fermilab) \cite{E774-Fermilab}, CHARM$^\dagger$ (CERN) \cite{CHARM-CERN-1,CHARM-CERN-2} and $\nu$-Cal$^\dagger$ (U70) \cite{Nucal}, and by astrophysical observations \cite{SN1987A}.
The middle region (zoomed in panel a) of Fig. \ref{fig:exclusionLimits}), expected to be covered by the projected sensitivities of HPS$^{\circ}$ (JLab) \cite{HPS} and LHCb$^{\circ}$ (CERN) \cite{LHCb-2,LHCb-1,LHCb-3,LHCb-4}, can be mostly accessed by experiments able to reconstruct the $A'$ decay vertex.

 \subsubsection{Bump hunting}
In these experiments the four-momentum of all the leptons from $A'$ decay is measured and the $\ell^{+}\ell^{-}$ invariant mass can be reconstructed. In the invariant mass spectrum, the hidden gauge boson will show up as a narrow peak over a smooth background, whose width is mainly determined by the experimental mass resolution, over a smooth  background. The mass resolution affects the sensitivity of such experiments, and firm control of systematic uncertainties is necessary since the $signal/background$ ratio can be as small as $\sim10^{-6}$. With reference to Fig. \ref{fig:exclusionLimits}a),
the exclusion region accessible to bump-hunting experiments is limited by the kinematic reach on the horizontal axis, and the background and the integrated luminosity on the vertical one, with the sensitivity (downwards) increasing with the luminosity. 
Fixed target experiments can achieve much higher intensities than  colliders (with the exception of flavor-factories) but the generally larger center-of-mass energy of the latter allow
the limits for $A^\prime$ masses to be extended to higher values. In some lucky cases, some of the background reactions can be suppressed at the trigger level and by proper offline cuts, exploiting the particular topology of the background processes, as for instance in the Bethe-Heitler case.  Unfortunately, this is usually not possible: 
for example, the kinematics of an $e^+e^-$ pair by an electron irradiated $A^\prime$  in a thin fixed target and of an $e^+e^-$ pair produced by virtual photon bremsstrahlung are identical, at the same $A'$ mass. The irreducible SM background represents the critical limit for a pure bump-hunt approach, therefore different techniques which use, for instance, passive shielding (beam-dump experiments) or further selections (displaced vertex reconstruction) need to be exploited to extend the coupling constant exploration to smaller values than those typically accessible. These different approaches will be described in the following paragraphs.
Different sorts of beams can be exploited to bump hunt purposes, as already mentioned at the beginning of this Section.
Fixed target experiments using electron beams and based on $A'$-strahlung production are APEX$^{\circ}$ (JLab) \cite{APEX1,APEX2}, A1, HPS, DarkLight$^{*}$ (JLab) \cite{DarkLight}, MAGIX$^{*}$ (MESA-Mainz) \cite{MAGIX}, NA64$^{\circ}$ (CERN) \cite{NA64-visible}. Fixed target experiments exploiting positron beams and based on the $e^{+}e^{-}$ annihilation in atoms are VEPP3$^{*}$ (BINP-Novosibirsk) \cite{VEPP3}, PADME$^{\circ}$ (LNF-Frascati) \cite{PADME-1,PADME-2}, MMAPS$^{*}$ (Cornell) \cite{MMAPS}. 
Potentially interesting results are awaited from Compton-like reactions induced by real photons in experiments like LEPS$^\ast$ (currently upgraded at SPring8) \cite{leps} and FOREST$^\ast$ (ELPH) \cite{forest},  able to measure also the recoiling electron, together with GlueX$^\circ$ \cite{gluex} and LEPS2$^\circ$ (SPring8) \cite{leps2}, which however can just measure either the di-lepton or the recoiling electron.
$e^{+}e^{-}$ collider experiments at relatively low center-of-mass energies are BABAR, BELLE-II$^\circ$ (KEK) \cite{BELLEII}, and KLOE$^\dagger$ (LNF-Frascati), for which the production modes include the $\phi \rightarrow \eta A'$ decay \cite{KLOE-1,KLOE-2,KLOE-3,KLOE-4}.  
Mu3e$^{*}$ (PSI) \cite{Mu3e-1,Mu3e-2} will be based on stopped muons measurements. Finally, experiments using hadron beams are NA48/2, HADES (GSI)$^\circ$ \cite{HADES}, LHCb.

\subsubsection{Beam-dump experiments for visible decays}
\label{BDvisible}
The data collected by beam-dump experiments performed in the Eighties  and Nineties for neutrino studies and to search for long-living new particles have been recently re-analyzed to set new constraints on the dark photon parameter space. 
The beam-dump  technique employs a high-intensity beam,  providing the large luminosity needed to probe the $A^\prime$ weak coupling, which gets completely absorbed by a thick target. A detector is placed tens or hundreds of meters downstream the dump after a passive shielding region of concrete and/or bedstone. If the dark photon is produced in the dump and lives long enough to cross both the dump and the shielding before decaying, its $e^{+}e^{-}$ or $\mu^{+}\mu^{-}$ decay products can be detected. Thus, beam-dump setups can be thought of as counting experiments where any evidence of an excess of di-lepton pairs not compatible with the expected background hints at the existence of new particles. \\ 
In case of electron beams the $A'$ can be produced via bremsstrahlung of primary and secondary electrons and by secondary positrons. The positrons of the electromagnetic showers generated in the dump can also produce $A'$ via non-resonant ($e^+e^- \rightarrow \gamma A'$) and resonant ($e^+e^- \rightarrow A'$) annihilation occurring on atomic electrons. Recently, it has been shown \cite{Marsicano1, Marsicano2} 
that showering effects in the dump affect both the $A'$ kinematics and the production yield,  and therefore they must be properly taken into account when extracting exclusion limits. In particular, the contribution of positron annihilation provides a larger sensitivity  in some kinematic regions with respect to the limits derived by the  $A'$-strahlung process. In proton dump experiments the $A'$ can be produced either directly, via proton or lepton $A'$-strahlung, or indirectly through meson decays. Leptons and mesons are secondary particles produced by proton scattering in the dump or in the electromagnetic and hadronic showers in the dump material.\\ 
The physics reach of these experiments depends on the downstream detector acceptance and on the total number of dumped particles. The higher the integrated luminosity, the larger the  exclusion region reported in Fig. \ref{fig:exclusionLimits}, which  extends towards smaller values of the coupling parameter. 
The use of high energy beams helps in terms of larger $A'$ production cross-sections and wider kinematics: the higher the energy of the primary electron or proton beam, the more forward boosted  the $A'$'s produced in the dump and their decay products. A typical feature of such experiments is that the physical reach strongly depends also on the  thickness of the shielding volume (dump included) and on the distance between the end of the shielding and the downstream detector, as the experiments are sensitive only to dark photons which decay in this region. On one side, with a decay path length of tens/hundreds meters, beam-dump experiments play as powerful tools to access the region of very low coupling parameters (corresponding to very long-lived $A'$), since the $A'$ decay length scales with $(\epsilon ^2 m_{A'})^{-1}$, as mentioned in Sec. \ref{decays}.  On the other side, this technique cannot probe ($\epsilon^2,\; m_{A'}$) parameters that correspond to too short $A'$ lifetimes, {\it i.e.} dark photons decaying too promptly in the dump or in the shielding. For this reason the sensitivity of beam-dump experiments is limited to the bottom-left regions of Fig. \ref{fig:exclusionLimits}, where $\epsilon$ and $m_{A'}$ are small enough to allow for a long decay path length.  
Electron beam dump experiments that have been re-analyzed in terms of $A'$ constraints are Konaka {\it et al.}$^\dagger$ (KEK) \cite{KEK}, E141, E137, E774, Davier {\it et al.}$^\dagger$ \cite{Orsay}.  All of these experiments have also the potentiality to deliver new information on Compton emission of dark photons \cite{compton}. Proton beam dump experiments are NA62$^{\circ}$ (CERN) \cite{NA62},  CHARM, PS191$^\dagger$(CERN) \cite{PS191},  NOMAD$^\dagger$ (CERN) \cite{NOMAD1,NOMAD2}, $\nu$-Cal, SHiP$^{*}$ (CERN) \cite{SHIP}, SeaQuest$^{\circ}$ (FNAL) \cite{Seaquest}. 

\subsubsection{Decay vertex reconstruction}
There is a region in the $(\epsilon^2, m_{A'})$ plane in which the $A'$ lifetime is too short to be explored by beam-dump experiments, but the coupling parameter  is so small that the \textit{signal/background} ratio is  too  unfavorable for a simple bump-hunt search to be effective. A third detection strategy, based on decay vertexes detection, can therefore be pursued to explore the parameter region where dark photons travel short but still detectable distances before decaying. A thin target is used to produce $A'$ via, for example, a brehmsstrahlung emission from an incoming electron, and the vertexes of the $\ell^{+}\ell^{-}$ pairs produced in the beam-target interaction can be reconstructed. The selection of lepton-pair events with a displaced vertex allows the background  from the prompt QED events (see Sec. \ref{QEDbackgrounds}) to be sensibly reduced, enhancing the experimental sensitivity in this region. This is the case, for example, of HPS \cite{HPS}. The experiment uses the CEBAF \cite{CEBAF} machine to accelerate electrons to energies between 1 and 6 GeV, which are then impinging on a thin tungsten target. The outgoing $e^{+}e^{-}$ pairs are detected in a compact, large acceptance forward detector consisting of a silicon vertex tracker and a lead tungstate electromagnetic calorimeter.
Another experiment applying this technique by the measurement of inclusive di-muon production is LHCb, that published the first exclusion plot obtained with the displaced vertex reconstruction approach \cite{LHCb-4}. The higher energy available as compared to HPS allows LHCb to explore a similar $\epsilon^2$ range at higher $A'$ masses. The experimental program of LHCb foresees to look for displaced vertices also in the $D^{*0} \rightarrow D^0 e^{+}e^{-}$ decay; this search requires, however, an upgrade of the current LHCb trigger system  and will start after the Long Shutdown 2 \cite{CERN-Report}.

\subsection{Invisible decay searches}
\begin{figure}[!ht]
\centering
\includegraphics[width=\linewidth]{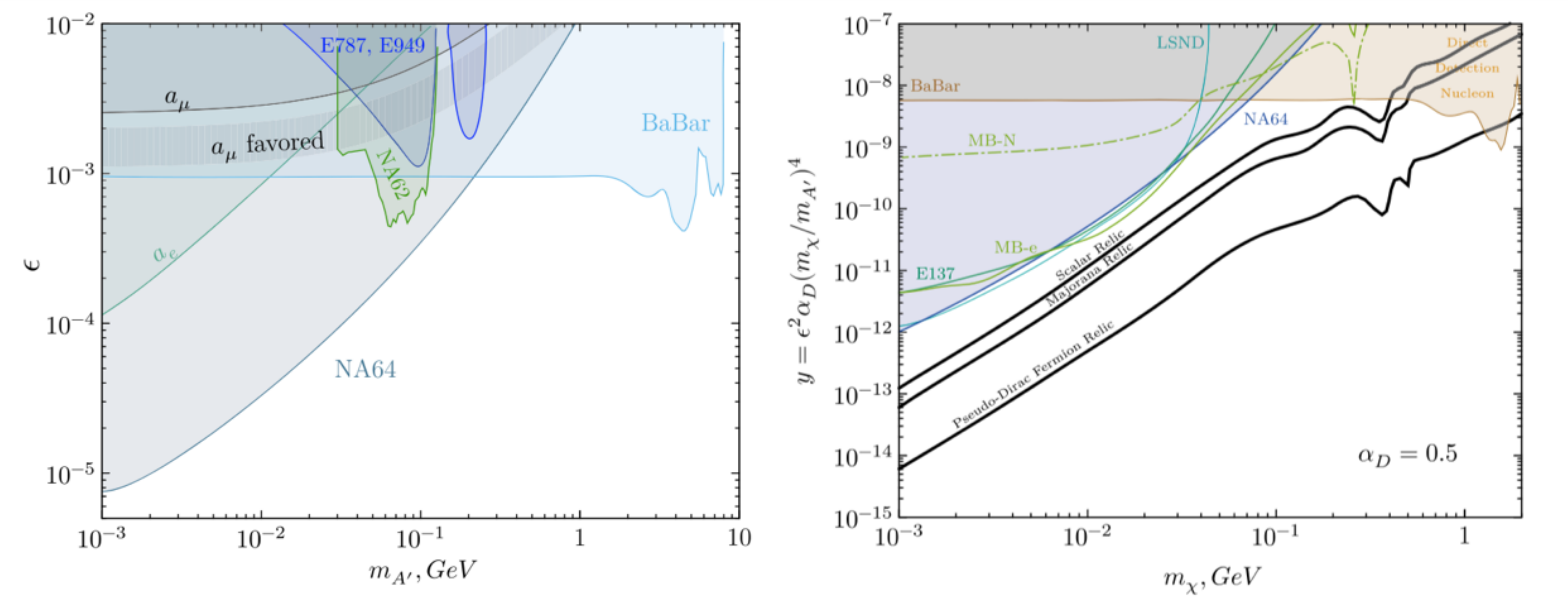} 
\put(-220,60){{\Large a)}}
\put(-40,60){{\Large b)}}
\\
\includegraphics[width=\linewidth]{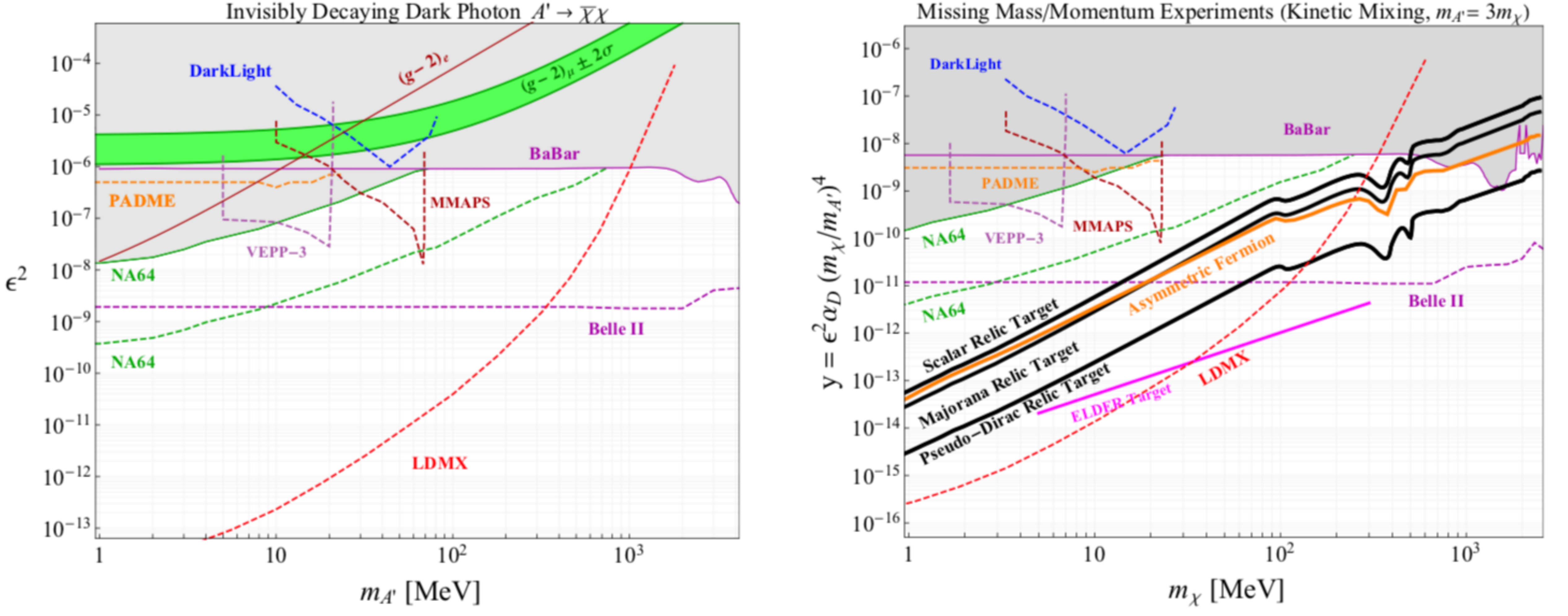}
\put(-220,40){{\Large c)}}
\put(-40,40){{\Large d)}}

\caption{Top panels: Present excluded regions at 90\% C.L., in the ($\epsilon,\; m_{A'}$) (a) and  ($y,\; m_{\chi}$) (with $\alpha_{D}=0.5$) (b) planes for dark photon invisible decay, from Ref.\cite{NA64-2019}. The favored parameter regions, which account for the observed relic DM density for the scalar, pseudo-Dirac, and Majorana LDM are indicated by the black solid lines. Additional constraints from E787 and E949 \cite{Davoudiasl, Essig-2013}, BABAR \cite{BaBar-invisible}, and NA62 \cite{NA62-invisible} experiments are shown together with the $a_{\mu}$ favoured region.  
Bottom panels: Sensitivity estimates (dashed lines) for various future proposed experiments based on missing mass/energy/momentum techniques (from Ref. \cite{US-cosmic-vision}): $(\epsilon^2, m_{A'})$ (c) and $(y,\; m_{\chi})$ (d) planes.}
\label{fig:exclusionLimits_invisible}
\end{figure}

As we have seen in Sec. \ref{properties}, the dark photon may couple to other Dark Sector particles,  charged under U(1)$_{D}$. The so-called \textit{invisible decay} searches are based on the general assumption that at least one new particle $\chi$ in the hidden sector has mass lower than $m_{A'}/2$, so that  the dominant $A'$ decay mode is invisible: $A' \rightarrow \chi \chi$, {\it i.e.} $\Gamma(A'\rightarrow\chi \chi)/\Gamma_{Tot}\approx 1$. The strength of this decay mode is controlled by the dark fine structure constant $\alpha_{D} = g_{D}^2/4\pi$. There is no {\it a priori} reason for the coupling of $A'$ to U(1)$_D$ particles to be suppressed, as it happens for the kinetic coupling to SM particles, so $\alpha_{D}$ is usually assumed to be $\mathcal{O}(1)$. 
The invisible decays can be detected by using missing-mass, missing-energy or missing-momentum experiments. Even the direct detection of the  DM particles from the decay can potentially be achieved via beam-dump experiments. 
Differently from the dark photon case, the search for LDM $\chi$ particles involves a  four-dimensional parameter space which includes the mass of the $\chi$ and of the $A^\prime$ and the two couplings $\alpha_{D}$ and $\epsilon$.
A further model dependence is associated with the spin and CP features of the DM candidate (which can be, for instance, a fermionic, scalar, Majorana or pseudo-Dirac particle). 
A summary of the latest constraints and sensitivity estimates is shown in the exclusion plots reported in Figs.\ref{fig:exclusionLimits_invisible},\ref{fig:exclusionLimits_invisible_dump}.
In particular, the top panels in Fig. \ref{fig:exclusionLimits_invisible} report the strongest limits in the $\epsilon\; vs\; m_{A'}$ (a) and $y=\epsilon^2\alpha_{D}(m_{\chi}/m_{A'})^4\; vs\; m_{A'}$ (b) plane. The dimensionless variable $y$  is convenient to quantify the sensitivity because it allows  clear limits to be fixed for each different CP LDM species (scalar, pseudo-Dirac, and Majorana type), consistently with the thermal paradigm in case of direct annihilation (black solid lines in Figs.\ref{fig:exclusionLimits_invisible}, \ref{fig:exclusionLimits_invisible_dump}). For each $m_\chi$ choice, in fact, there is a unique value of $y$ compatible with thermal freeze-out, independent of the values of $\alpha_{D}$, $\epsilon$, and $m_{\chi}/m_{A'}$ \cite{US-cosmic-vision}: 

\begin{equation}
    \label{eq:y}
    \sigma v (\, \chi\chi \rightarrow A'^{*} \rightarrow \mathrm{SM}\;\mathrm{SM}) \propto 
    \epsilon^2\alpha_{D} \frac{m^2_{\chi}}{m^4_{A'}}=\frac{y}{m^2_{\chi}}
\end{equation}
The bottom panels (c, d) of Fig. \ref{fig:exclusionLimits_invisible} include sensitivity estimates of several experiments proposed for the future. 
When applicable, all bounds and projections assume the conservative prescriptions $m_{A'} = 3 m_{\chi}$ and $\alpha_{D}$=0.5. For smaller values of $\alpha_{D}$ and $m_{\chi}/m_{A'}$, the limits shift downwards and a more extended region of the parameter space is covered; close to the resonance region, $m_{A'} \sim 2m_{\chi}$, the estimated limits might depart from Eq.\ref{eq:y}; for larger values of the $m_{\chi}/m_{A'}$ ratios, {\it i.e.} when $m_{A'} < 2m_{\chi}$, the DM annihilation proceeds through $\chi\chi\rightarrow A'A'$, which is independent of the  $\epsilon$ SM coupling.  
Fig. \ref{fig:exclusionLimits_invisible_dump}a) shows the limits for leptophilic DM, in which  the DM couples preferentially to leptonic currents, whereas limits for the  scenario in which the DM couples preferentially to the barionic currents (leptophobic DM) are reported in Fig. \ref{fig:exclusionLimits_invisible_dump}b). Some details on these two models can be found, for example, in Refs.\cite{darksectors16,US-cosmic-vision}. Fig. \ref{fig:exclusionLimits_invisible_dump}c) shows new limits extracted from a recent re-analysis of some electron beam dump experiments \cite{Marsicano1,Marsicano2} where the $A'$ production from positron annihilation in the dump has been included for the first time.   
Finally, limits related to the inelastic DM scenario (see Sec. \ref{decays}), which includes two $\chi_1$ and $\chi_2$ DM mass states, are reported in Fig. \ref{fig:exclusionLimits_invisible_dump}d) \cite{izaguirre-2017}.

\subsubsection{Missing mass}
This approach requires the reconstruction of all (but one at most) the SM final state particles and a well-known initial state: a DM candidate is identified as a resonant signal over a smooth background in the missing mass distribution of the measured final state system. \\
The DM particles can for instance be produced in exclusive reactions such as $e^{+}e^{-} \rightarrow \gamma A',\; A'\rightarrow\chi{\chi}$ using an electron-positron collider as in BABAR or BELLE-II. Indeed, the strongest constraints to-date on the parameter space for DM masses $m_{\chi} \gtrsim 100$ MeV have been derived from the BABAR mono-photon searches \cite{BaBar-invisible}. Here the resolution on the $\chi$ mass is dominated by the photon energy resolution, which decreases monotonically from $\sigma(M_{\chi}^{2})=1.5$ GeV$^2$ for $m_{A'}\sim 0$ to $\sigma(M_{\chi}^{2})=0.7$ GeV$^2$ at $m_{A'}\sim 8$ GeV. \\
Another possibility based on $e^+e^-$ annihilations is to use a positron beam impinging on fixed target, as in VEPP3,  PADME and MMAPS. The real photon energy and direction is measured by an electromagnetic calorimeter, and a single-photon event is searched for.
In all the cases a signal of the dark photon would appear as a narrow peak, centered at $m_{A'}$, over a continuous background distribution  generated by all the events with a single high-energy photon and no other signal in the final state. This of course requires a large detector hermeticity and full control of all the possible spurious accidental signals in the experimental setup.\\
Quite recently NA62 \cite{NA62-invisible} published new results on the search for an invisible $A'$  decay applying a missing-mass technique to the full reconstruction of the $K^{+} \rightarrow \pi^{+} \pi^{0}, \pi^{0} \rightarrow A' \gamma$ decay chain.
Secondary positive kaons are produced by the interaction of a primary 400 GeV/$c$ proton beam on a beryllium target and identified by a differential Cherenkov counter. The kaon is tracked by silicon pixel detectors while the charged pion momenta are measured by a straw-based magnetic spectrometer ($\sigma_{p}/p=(0.3-0.4)\%$); the recoiling photon is detected in a liquid krypton electromagnetic calorimeter (with resolution along the transverse coordinate $\sim 1$ mm, and $\sigma_{E}/E\sim4.8\%/\sqrt{E[\mathrm{GeV}]}$). The missing mass is finally reconstructed following the assumption that the photon is emitted from the $K^+$ decay vertex (with a resolution $\sigma(M^{2}_{Miss})/M^{2}_{Miss}\sim (5.7-6.5)\%$ for $A'$  with mass in the range $m_{A'}=30-130$ MeV.) 
Additional coverage in the same parameter region (see Fig. \ref{fig:exclusionLimits_invisible}a) is provided  \cite{Davoudiasl, Essig-2013} by the data of E787$^\dagger$ \cite{e787_04} and E949$^\dagger$ \cite{e949_09} BNL experiments, both dedicated to rare $K^+$ decays studies. 
Another production mechanism was proposed for  DarkLight, based on the  $e^{-}p \rightarrow e^{-}p A', A'\rightarrow\chi{\chi}$ reaction. DarkLight is expected to start operations soon at the JLab Low Energy Recirculating Facility (LERF) \cite{LERF}, with the beam impinging on a gaseous hydrogen target \cite{DarkLight}: it will be able to provide a complete reconstruction of the full final state kinematics of the production reaction.\\
Factors limiting the sensitivity of missing-mass experiments are the available luminosity, the momentum resolution and the hermeticity of the experimental setup, which is especially critical to keep under control the background from processes in which the emitted particles leave the apparatus undetected, mimicking an $A'$ invisible decay. 
The momentum resolution can be of course spoilt by several effects, the most sizeable of which is the multiple Coulomb scattering suffered by charged particles in tracking detector materials and volumes, and the uncertainty in the impact point position and in the energy released by photons in electromagnetic calorimeters. 
Finally, in mono-photon final state searches the main background source is given by QED continuum reactions such as $e^{+}e^{-} \rightarrow \gamma \gamma$, with $\gamma$'s emitted at large-angle, or $e^{+}e^{-} \rightarrow \gamma e^{+}e^{-}$,  with the leptons escaping the apparatus acceptance. 

\subsubsection{Missing energy/momentum}
When the dark mediator is produced in fixed-target reactions on $Z$ atomic mass number nuclei, using for instance electron beams as in $e^{-}Z \rightarrow e^{-}Z A'$, its invisible decay can be inferred through the missing energy/momentum that it carries away. 
The main challenge of this approach is an highly performing background rejection, which relies strongly on the detector hermeticity and, in most of the  cases, on the exact knowledge of the initial and final state kinematics.\\
In a missing-energy experiment, the crucial point is the balance between the beam energy and the energy of all the final state particles. The expected signal yield can be enhanced by using a thick active target, as a sort of active beam-dump, where the $A'$ is produced and the energy of the recoil electron is measured. 
This is the case of NA64 \cite{NA64-2019} where the occurrence of $A' \rightarrow invisible$  decay would provide an excess of events featuring a single shower in a first electromagnetic calorimeter (ECAL), used as an active target, and a negligible energy release in the downstream part of the detector (a second, hadronic, calorimeter HCAL, plus veto detectors).  Candidate events are requested to have $E_{HCAL} < 1$ GeV and a missing energy $E_{miss} = E_{0} - E_{ECAL} > 50$  GeV, with a $E_0 = 100$ GeV  electron beam.
The hermeticity of the ECAL in containing the full shower is indeed crucial to suppress the background determined by detection inefficiencies. To this purpose the Shashlik ECAL used in NA64 has a total thickness of 40 $X_{0}$ (radiation lengths), in which wavelength shifter fibers are spiral-wise interleaved in order to suppress the energy leaks. The energy resolution of the
ECAL is $\sigma_{E}/E \sim 9\%/\sqrt{E[\mathrm{GeV}]}$, and the longitudinal and lateral segmentation allows the shower profile reconstruction for further hadronic background rejection.
Another peculiar feature of this experiment is the detection of synchrotron radiation from the beam electrons, used as well to suppress the background deriving from the hadron contamination in the beam. 
Using this technique, NA64  recently published  the most stringent constraints on the $A'$-photons mixing strength, as well as  exclusion plots for  scalar and fermionic DM in the mass range below 0.2 GeV. A possible future NA64$^{++}$ experiment using high-energy electron, muon and hadron beams and an upgraded detector to cope with  higher beam intensities has recently been proposed \cite{CERN-Report}. \\
In a missing momentum experiment, as LDMX$^\ast$ \cite{LDMX} proposed to run at the SLAC DASEL facility on LCLS-II \cite{LCLSII} or at CERN (e-SPS), the momentum of each incoming electron  must be measured together with the energy and momentum of the outgoing particles. Contrarily to a typical missing-energy experiment, this approach clearly requires the use of a thin target.
In LDMX a Si-based tracking systems immersed in a dipole field is used to track the electron before it impinges on the target. Downstream, a second Si-based tracking system in a magnetic field, and two hermetic (electromagnetic and hadronic) calorimeters measure the momentum and energy of the produced particles. The experimental signature for an $A'$ invisible decay consists of a soft electron scattered at wide angle with no other particles in the final state.  The selection of a low-energy and wide-angle recoil electron allows efficient background rejection and signal selection. Indeed, as long as the $A'$ or the $\chi{\chi}$ pair is heavy as compared  to the electron mass, the differential cross-section for DM production is peaked when the DM carries away most of the beam energy, and the electron a relatively small part of it. These kinematic features are opposite to what occurs in ordinary bremsstrahlung events. 
The ability to separate the signal from background using the recoil transverse momentum is generally limited by the multiple scattering in the target.  For 4 GeV electrons impinging on a 10\% $X_{0}$ target, 
LDMX simulations \cite{LDMX} show an average transverse momentum uncertainty for the recoiling electron ($E_{recoil}$ from $\sim$0.1 to $\sim$2 GeV) on the order of 4 MeV, that is the typical smearing due to the multiple scattering in the target.
The missing momentum approach has several advantages compared to other techniques: in fact, the missing mass/energy approaches generally suffer from higher background levels due to the fewer kinematic constraints, while beam-dump experiments are penalized by the additional interaction of the DM in the detector materials.
On the other hand, in order to reach the full potential of the missing-momentum technique,  demanding constraints must be respected by the setup and beam-line equipment: for instance, an
extremely low electron density per beam bunch (1--5 $e^{-}$/bunch) is required to allow single electron tracking  before the target; this low-current requirement must be compensated by high repetition rates ($\sim 10^{9}$ Electrons-On-Target (EOT)/s) in order to reach the necessary integrated luminosity ($10^{14}-10^{16}$ EOT).
These figures result in an effective running period of around 100 days, which typically corresponds to some years of data taking on floor.
Moreover, a large beam spot ($\sim$10 cm$^{2}$) is needed to ease the identification of single electrons, as well as low spread-out occupancy and radiation doses. Finally, fast and hermetic detectors must resolve the energies and angles of every incident and scattered electron, simultaneously rejecting a large variety of potential background processes whose rate spans over many orders of magnitude.     

\begin{figure}[!ht]
\centering
\includegraphics[width=\linewidth]{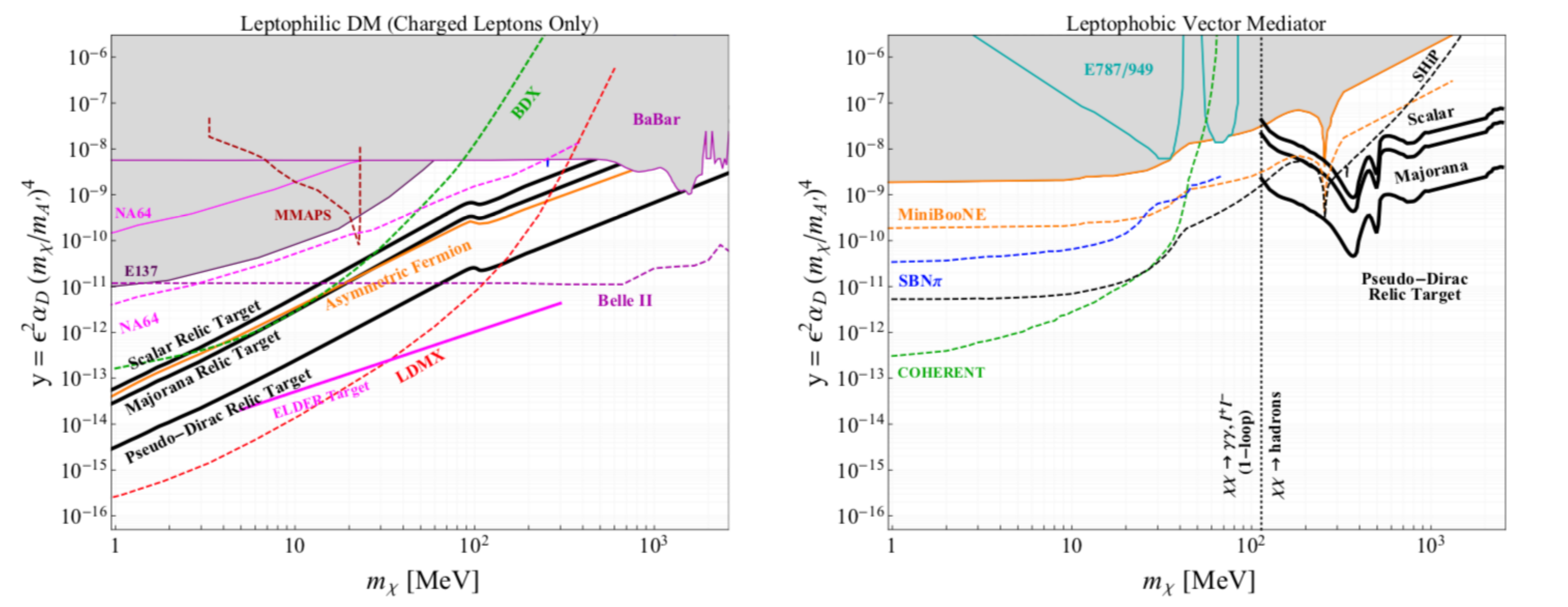}
\put(-220,40){{\Large a)}}
\put(-40,40){{\Large b)}}
\\
\includegraphics[width=\linewidth]{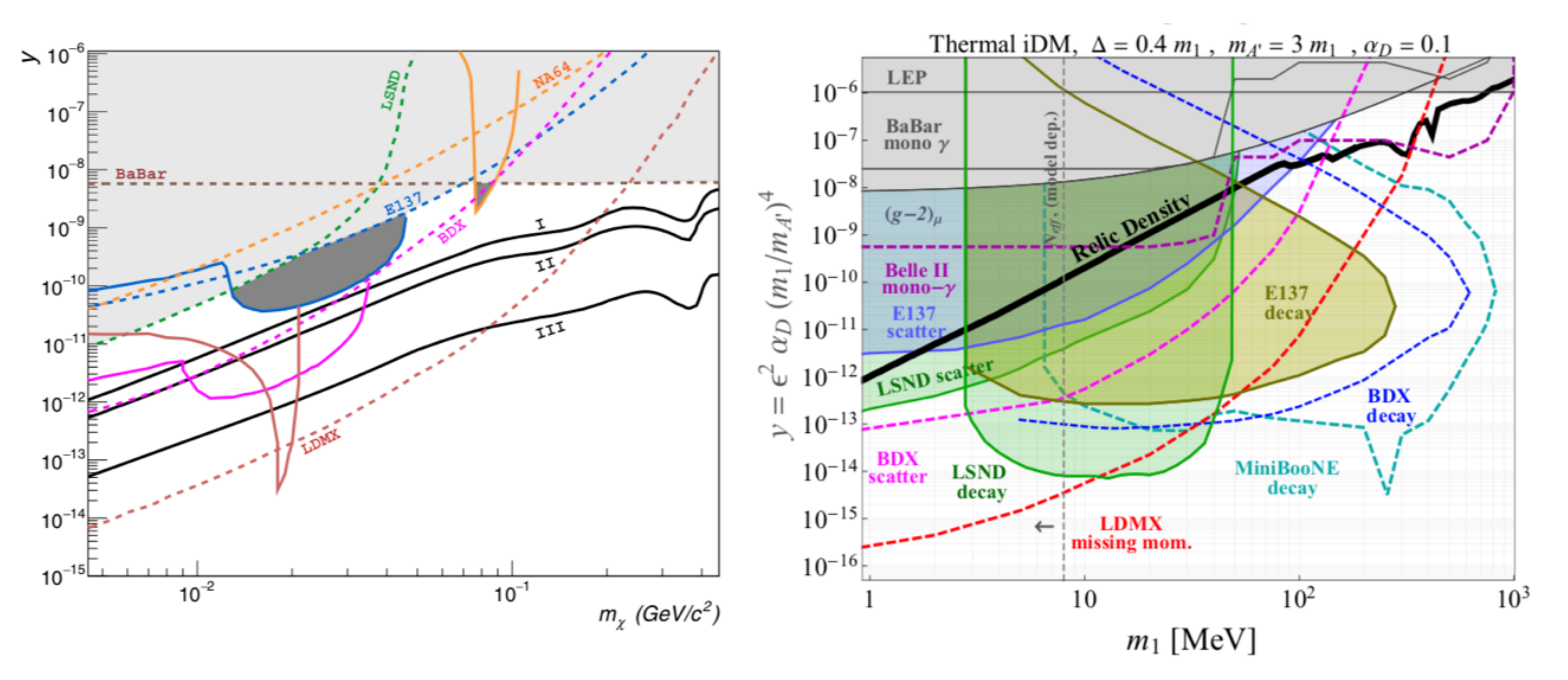}
\put(-220,30){{\Large c)}}
\put(-40,30){{\Large d)}}
\caption{Top panels: Limits in $(y, m_{\chi})$ parameter space for  electron (leptophilic, a)) and nucleon (leptophobic, b)) couplings. The expected sensitivities of various proposed beam-dump experiments are reported (from Ref. \cite{US-cosmic-vision}). Fig. c): $(y,\; m_\chi)$ exclusion plane for lower masses, leptophilic coupling; the dashed and continuous lines show the exclusion limits obtained by considering $A'$-strahlung and positron annihilation only, respectively (from Ref. \cite{Marsicano2}). Fig. d): existing bounds (filled areas) and sensitivities of proposed experiments (dashed lines) for inelastic LDM $\chi_1$ detection (from Ref. \cite{izaguirre-2017}). For this DM scenario beam-dump experiments can be more effective than missing mass experiments at higher energies (see text for details).   
}
\label{fig:exclusionLimits_invisible_dump}
\end{figure}

\subsubsection{Beam-dump experiments for invisible decays}
\label{BDinvisible}
In this approach the $A^\prime$ invisible decay is probed by directly detecting the DM particles through their inelastic collisions. There are many similarities with the dump techniques for the visible decay channel described in Sec. \ref{BDvisible}, as the same beam-dump experiment may also be able to explore both decay scenarios as in the case, for example, of E137. 
In both visible and invisible decay searches, the mechanism for the $A'$ production in the dump is clearly the same, thus the inclusion of showering  and production by $e^{+}e^{-}$ annihilation  in the dump has sizeable effects also in the sensitivity of invisible decay searches \cite{Marsicano2} (see Fig. \ref{fig:exclusionLimits_invisible_dump}c). The $A'$ decay products, either visible or invisible, carry out most of the beam energy and are very forward boosted. Therefore a downstream detector with a relatively compact transverse section, as an electromagnetic calorimeter placed behind some passive shielding, can be used to detect DM particles.
An important difference between visible and invisible searches at dump experiments lies in the sensitivity to different $A'$ lifetimes. While for visible decays a beam-dump experiment is sensitive to long-lived $A'$ only,
in the invisible decay case there is not such a restriction. Therefore, the $A'$ can decay anywhere along its flight path to the detector, even promptly in the dump (as a $\mathcal{O}(1)$ $\alpha_{D}$ would require), since the produced DM particles can cross the dump and the shielding volume undisturbed.\\
If compared to missing mass/energy/momentum experiments, the beam-dump approach offers the advantage of probing the DM interaction twice, providing direct sensitivity to the dark sector mediator coupling $\alpha_{D}$. The price to pay for the detection of a DM scattering reaction is a huge reduction of the expected signal yield. The DM detection at a downstream detector is achieved through the $\chi e^{-}$ or $\chi N$ scattering processes, which imply an additional $A'$ exchange. Therefore, the DM signal yields scale as $\epsilon^{4}\alpha_{D}$, in which an  $\epsilon^{2}\alpha_{D}$ factor comes from DM production vertex, and an additional $\epsilon^{2}$ comes from the DM interaction, to be compared to $N_{signal}\propto\epsilon^{2}$ typical of missing mass/energy/momentum experiments. 
To (partially) compensate for the small scattering probability a large proton/electron beam flux is required --as a rule of thumb, new beam-dump experiments must collect more than $10^{20}$ electrons/protons on target to explore virgin regions of the parameter space. 
Beam-dump techniques are particularly suited for the detection of iDM particles (see Sec. \ref{richSectors} and \ref{decays}), for which they can offer superior sensitivity as compared to missing mass/energy/momentum experiments, despite the most favourable signal yields available in the latter. In fact, in a beam-dump experiment, the $\chi_{1}$ produced in the dump can upscatter off the detector materials to generate a $\chi_{2}$ state (Fig. \ref{fig:iDM}b)). The projected sensitivities of beam-dump experiments able to detect the SM recoil products of the $\chi_1$ scattering  are labelled as ``scatter" in Fig. \ref{fig:exclusionLimits_invisible_dump}d). If the $\chi_2$ state de-excites
($\chi_2\to A^{\prime\ast}\chi_1$) inside the detector (Fig. \ref{fig:iDM}b)), this would lead to an even more striking signal:  one recoil-target body (an electron, nucleon or nucleus)  plus an $e^{+}e^{-}$ pair from the $A'^{\ast}$ decay. The exclusions limits for this scenario for some beam-dump experiments are indicated as ``decay" in Fig. \ref{fig:exclusionLimits_invisible_dump}d). On the contrary, the de-excitation of the $\chi_{2}$ inside the active target in a missing energy experiment would mimic the most critical background source, {\it i.e.} bremsstrahlung events with a photon converting to $e^{+}e^{-}$, and the detection of such an effect would be prevented.
Generally speaking, the limiting factor on the sensitivity of beam-dump experiments comes from beam-related neutrinos. The DM signature in the detector is similar to that of neutrino interactions (except for the aforementioned iDM case), nevertheless topological and kinematic selections can be used to reduce the neutrino background contribution.
This is also the reason why the strongest existing constraints for dump experiments come from neutrino experiments as LSND$^\circ$ (Los Alamos) \cite{LSND} and MiniBooNE$^\circ$ (FNAL) \cite{MiniBoone}, which complement the re-analysis of the E137  electron beam dump data. Such  limits dominate over the BABAR constraints for lower mass DM ($m_{\chi} \lesssim 100$ MeV).\\ 
Other proton beam-dump experiments which, besides LSND  and MiniBooNE, can potentially exclude unexplored regions of the parameter space are SHiP, Coherent$^{*}$ (ORNL) \cite{Coherent} and SBN$^{*}$ (FNAL) \cite{SBN}. 
All of them are  multipurpose experiments with a wide neutrino research program. \\
A frontier experiment specifically designed and optimized to search for light DM by dumping an intense electron beam is, finally, BDX$^{*}$ (JLab) \cite{BDX1,BDX2,BDX3}. The experiment is expected to run in a dedicated underground facility  located $\sim$20 m downstream of the CEBAF Hall A beam-dump. It will use a 10.6 GeV $e^-$ beam provided by the CEBAF machine and is expected to collect up to $10^{22}$ EOT. The detector will consist of two main components: a CsI(Tl) electromagnetic calorimeter (ECAL) and a veto system for the background rejection. The expected signature of the DM interaction in the ECAL is a $\sim$GeV electromagnetic shower matching with the absence of activity in the surrounding active veto counters. A proof of concept setup has recently been installed at JLab in a simplified unshielded configuration. It is presently using, in parasitic mode, a 2.2 GeV $e^-$ beam and is expected to run for one year. A small and compact prototype detector, called BDX-MINI, is composed of a PbWO$_4$ electromagnetic calorimeter, surrounded by a layer of tungsten shielding and two hermetic plastic scintillator veto systems. This early stage prototype represents a first dedicated new-generation beam-dump experiment, whose physics reach will cover a kinematic region as large as what has been excluded so far by all the non-dedicated previous experiments. 

\section{Conclusions}

This Review  reports a limited length overview of the presently lively and sparkling experimental activity, supported by a a likewise excitement on the theoretical side, in the search of hidden-sector Dark Matter and, in particular, of the mediator of a still unknown new interaction, the dark photon. 
This search is well motivated among all possible scenarios formulated in Dark Matter science  
as, based on the hypothesis of a relatively small mass in the sub-GeV range, the dark photon discovery is in principle accessible to quite a sizeable number of small-scale experiments currently on floor, under construction or in development in many laboratories all over the world. They are based on a remarkable variety of different experimental techniques, and most of them are sharply targeted to provide high sensitivity and precision results in well defined and so far unconstrained regions of the DM parameter space; as such, all of them are highly complementary and synergistic.
The search for the dark photon is a typical example of a fully transverse investigation. On one side, in fact, it requires a thorough integration of all the collected observations by means of a solid theoretical framework, able to harmonize many inputs coming from different experimental fields involving not only particle physics and quantum mechanics  related-effects, but also astrophysics, cosmology and thermodynamics. On the other hand, the comprehensive approach calls for a collective effort aimed to provide the most complete information as possible. This includes inputs from already finished experiments whose data can relive in dedicated novel re-analysis, or from experiments primarily dedicated to other purposes, like neutrino physics. All contributions will be valuable and necessary, from small scale direct search experiments to medium scale setups at accelerators, and to big experimental facilities at the largest existing to-date colliders. All of them are and will be useful to provide important tiles in the biggest puzzle of today's particle physics and cosmology, the identification of Dark Matter and the understanding of its nature and behaviour.
The endeavour is indeed very ambitious, but the importance of the target is well worth all the possible efforts.

\bibliographystyle{model1-num-names}
\bibliography{references}


\end{document}